\documentclass[final,5p,times,twocolumn]{elsarticle}

\usepackage{amsmath}
\usepackage{amssymb}
\usepackage{amsfonts}
\usepackage{graphicx} 
\usepackage{dcolumn} 
\usepackage{bm} 
\usepackage{hyperref} 
\hypersetup{colorlinks=true,citecolor=blue,urlcolor=blue,linkcolor=blue}

\usepackage{url}
\usepackage{xcolor}
\definecolor{newcolor}{rgb}{.8,.349,.1}

\providecommand{\abs}[1]{\lvert#1\rvert}

\newcommand{\vect}[1]{\vec{#1}}
\newcommand{\hatvect}[1]{\hat{\!\vec{\,#1}}{}}
\newcommand{\mat}[1]{\mathbf{\boldsymbol{#1}}}
\newcommand{\vectmat}[1]{\!\vec{\,\mat{#1}}}

\journal{Journal of Magnetism and Magnetic Materials}

\begin{document}


\begin{frontmatter}

\title{Determining Perpendicular Magnetic Anisotropy in Fe/MgO/Fe Magnetic Tunnel Junction: A DFT-Based Spin-Orbit Torque Method}

\author[ncuphy]{Bao-Huei Huang\corref{cor1}}
\ead{lise811020@gmail.com}
\author[ntuphy,sinica]{Yu-Hsiang Fu}
\author[sinica]{Chao-Cheng Kaun}
\author[ncuphy]{Yu-Hui Tang\corref{cor1}}
\ead{yhtang@cc.ncu.edu.tw}

\cortext[cor1]{Corresponding author}

\affiliation[ncuphy]{
	organization={Department of Physics, National Central University},
	city={Taoyuan},
	postcode={32001},
	country={Taiwan}}
\affiliation[ntuphy]{
	organization={Department of Physics, National Taiwan University},
	city={Taipei},
	postcode={10617},
	country={Taiwan}}
\affiliation[sinica]{
	organization={Research Center for Applied Sciences, Academia Sinica},
	city={Taipei},
	postcode={11529},
	country={Taiwan}}

\begin{abstract}
	In our \textsc{JunPy} package, we have combined the first-principles calculated self-consistent Hamiltonian with divide-and-conquer technique to successfully determine the magnetic anisotropy (MA) in an Fe/MgO/Fe magnetic tunnel junction (MTJ). We propose a comprehensive analytical derivation to clarify the crucial roles of spin-orbit coupling that mediates the exchange and spin-orbit components of spin torque, and the kinetic and spin-orbit components of spin current accumulation. The angular dependence of cumulative spin-orbit torque (SOT) indicates a uniaxial MA corresponding to the out-of-plane rotations of magnetic moments of the free Fe layers. Different from the conventional MA energy calculation and the phenomenological theory for a whole MTJ, our results provide insight into the orbital-resolved SOT for atomistic spin dynamics simulation in emergency complex magnetic heterojunctions.
\end{abstract}




\end{frontmatter}


\section{Introduction}

A perpendicular magnetic anisotropy (PMA) based magnetic tunnel junction (MTJ) is proposed as a next-generation memory device for spintronics because of its high speed in writing and reading processes and low energy consumption \cite{Ikeda2010}.
From the theoretical point of view, the spin-spin coupling (SSC) and the spin-orbit coupling (SOC) dominate the fundamental mechanism of spintronics. The exchange interaction due to SSC between two spins keeps the spins parallel or anti-parallel and results in the magnetism of materials.
For electrons in solids, the atomic orbitals couple to the lattice and produce bonding among them. The SOC couples the spin direction to a specified axis along with the bonding and results in the magnetic anisotropy (MA) effect.
The corresponding magnetic anisotropy energy (MAE) describes the strength to rotate the spin direction away from the easy axis due to SOC.

For the MAE calculation, the energy method is usually the way to estimate the value by computing the total energy in different magnetization angles, and the energy difference between the maximum and minimum could be the MAE by force theorem \cite{Weinert1985,Daalderop1990}, which leads to several first-principles calculations have been performed to study the interfacial PMA effect \cite{Peng2015,Peng2017}. On the other hand, the spin-orbit torque (SOT) method is proposed to estimate the MAE because the total energy calculation would be difficult due to the numerical stability and high precision requirement during the self-consistent field (SCF) process \cite{Wang1996,Pick1999}. The difficulty of the torque method emerges on the calculation of magnetic noncollinear system, for which the SCF process is usually more time-consuming than that of the collinear system due to additional degrees of freedom on the other spin directions. Thanks to the development of high performance computing, the noncollinear calculations are achievable recently.

The combination of SSC/SOC and nonequilibrium transport under bias voltages (currents) generates the spin accumulation and controls the magnetization switching. For example, the current-driven spin-transfer torque (STT) is based on the spin-transfer between the spin-polarized current and the magnetization via the exchange interaction due to SSC \cite{Locatelli2013}. For the current-driven SOT, an in-plane spin-polarized current flowing through a heavy metal suffers from an effective magnetic field produced by SOC, and exchanges the spin as well as the orbital angular momentum with the magnetization of the free layer at the interface \cite{Manchon2019}.

In this study, we propose the calculations of current-driven STT and interfacial SOT in a framework of density functional theory (DFT).
Our developed \textsc{JunPy} package \cite{JunPy,Huang2021,Tang2021} combined first-principles calculations with nonequilibrium Green's function (NEGF) formalism achieves general and universal calculations for spin torque as well as spin current accumulation.
We first present the theoretical formalism to derive the spin torque and spin current accumulation in the consideration of SOC, and clarify the influence on the local moment dynamics described by the Landau-Lifshitz-Gilbert (LLG) equation.
Next, first-principles calculations on an Fe/MgO/Fe MTJ in noncollinear magnetic configurations are performed to compute the current-driven STT by applying electrical bias across the junction and the interfacial SOT by considering SOC in the SCF process.
Finally, the spin torque and spin current accumulation are calculated by our \textsc{JunPy} package, in which the electronic structure is self-consistently obtained by the DFT-NEGF calculation, to study the current-driven STT and the interfacial magnetic anisotropy by SOT.

\section{Theoretical formalism}

\subsection{System Hamiltonian operator}

For a noncollinear spin-polarized system in a framework of density functional theory (DFT), where the ground state energy of a system can be uniquely expressed as a functional of the ground state electronic density, the effective single-electron Hamiltonian operator for a specified orbital $\lambda$ can be defined as a sum of the kinetic energy, spin-independent effective potential operator $\hat{\mathcal{V}}_{\lambda}^{0,\mathrm{eff}}$, exchange and correlation potential operator $\hat{\mathcal{H}}^{\mathrm{XC}}$, and spin-orbit potential operator $\hat{\mathcal{H}}^{\mathrm{SO}}$ \cite{Zhang2004,Ralph2008,Manchon2009,Haney2010}:
\begin{equation}\label{H_all}
	\hat{\mathcal{H}}=\frac{\hatvect{p}^{\,2}}{2m_e}+\hat{\mathcal{V}}_{\lambda}^{0,\mathrm{eff}}+\hat{\mathcal{H}}^{\mathrm{XC}}+\hat{\mathcal{H}}^{\mathrm{SO}}~,
\end{equation}
where $\hatvect{p}$ is the momentum operator and $m_e$ is the electron mass.

For magnetic materials with a local magnetic moment $\vect{M}_\lambda$, the unbalanced spin-up and spin-down density of states provide an exchange field $\vect{B}_{\lambda}^{\mathrm{XC}}=\mu_0\vect{M}_{\lambda}$ to couple with the spin angular momentum operator $\hatvect{S} = (\hat{S}^x,\hat{S}^y,\hat{S}^z)$:
\begin{equation}
	\hat{\mathcal{H}}^{\mathrm{XC}} = \gamma \hatvect{S}\cdot \vect{B}^{\mathrm{XC}}_\lambda = \xi_{\lambda}^{\mathrm{XC}}\hatvect{S}\cdot \vect{M}_\lambda~,
\end{equation}
where $\xi_{\lambda}^{\mathrm{XC}}$ represents an effective coupling strength and $\gamma$ is the electron gyromagnetic ratio. Meanwhile, according to the special theory of relativity, an electron moving inside an atom will see an effective magnetic field $\hatvect{B}^{\mathrm{SO}}_{\lambda}$ in its own reference frame. The so-called spin-orbit coupling contributes to the spin-orbit potential operator:
\begin{equation}
	\hat{\mathcal{H}}^{\mathrm{SO}} = \gamma \hatvect{S} \cdot \hatvect{B}^{\mathrm{SO}}_\lambda = \xi_{\lambda}^{\mathrm{SO}}\hatvect{S} \cdot \hatvect{L}_\lambda~,
\end{equation}
where $\hatvect{L}_\lambda = \hatvect{r}_\lambda \times \hatvect{p}$ with $\hatvect{r}_\lambda$ being a position operator and $\xi^{\mathrm{SO}}_\lambda$ is an effective SOC strength value.

\subsection{Spin continuity equation}
\label{sec:spin_continuity}

We start from the spin continuity equation to derive the spin torque and spin current accumulation. Introducing $\hat{P}_{\lambda}$ and $\hat{P}_{\lambda}^{\dagger}$ as the projection operators to the $\lambda$ orbital, the spin angular momentum operator of the $\lambda$ orbital can be defined as
\begin{equation}
	\hatvect{S}_{\lambda}=\frac{1}{2}\left( \hat{P}_{\lambda}^{\dagger}\hatvect{S}+\hatvect{S}\hat{P}_{\lambda} \right),
\end{equation}
and the expectation value, $\vec{S}_\lambda = \langle\hatvect{S}_\lambda\rangle$, is the spin density at the $\lambda$ orbital.
Here $\sum\nolimits_{\lambda}^{}{\hat{P}_{\lambda}}=\sum\nolimits_{\lambda}^{}{\hat{P}_{\lambda}^{\dagger}}=\hat{I}$ satisfies the completeness relation with $\hat{I}$ being the identity operator.
Note that $\hat{P}_{\lambda}$ and $\hat{P}_{\lambda}^{\dagger}$ are not identical for nonorthogonal basis used in our first-principles implementation of Sec.~\ref{sec:implementation}.

Based on the Heisenberg equation of motion, we have a continuity equation of the total spin density projected to a specified $\lambda$ orbital in the form of
\begin{equation}\label{dsdt_1}
	\frac{d\vec{S}_{\lambda}}{dt} = \frac{1}{i\hbar} \left<\left[ \hatvect{S}_{\lambda} ,\hat{\mathcal{H}} \right]\right> =  \vec{T}_{\lambda}^{\,\mathrm{S}} + \vec{\Phi}_{\lambda}^{\mathrm{S}}~.
\end{equation}
The first term is the total spin torque acting on the $\lambda$ orbital \cite{Haney2007,Kalitsov2006}:
\begin{equation}\label{spin_torque_operator_1}
	\vec{T}_{\lambda}^{\,\mathrm{S}}= \frac{1}{2} \left<\left( \hat{P}_{\lambda}^{\dagger}\hatvect{T}^{\mathrm{S}}+\hatvect{T}^{\mathrm{S}}\hat{P}_{\lambda} \right)\right>~,
\end{equation}
where
\begin{equation}\label{spin_torque_operator_2}
	\hatvect{T}^{\mathrm{S}}=\frac{1}{i\hbar}\left[ \hatvect{S},\hat{\mathcal{H}} \right]
\end{equation}
is the total spin torque operator. The second term is the spin current accumulation at the $\lambda$ orbital:
\begin{equation}\label{spin_current_operator_1}
	\vec{\Phi}_{\lambda}^{\mathrm{S}} = \frac{1}{2} \left<\left( \hatvect{S}\hat{\Phi}_{\lambda}+\hat{\Phi}_{\lambda}^{\dagger}\hatvect{S} \right)\right>~,
\end{equation}
where
\begin{equation}\label{spin_current_operator_2}
	\hat{\Phi}_{\lambda}=\frac{1}{i\hbar}\left[ \hat{P}_{\lambda},\hat{\mathcal{H}} \right] ~;~ \hat{\Phi}_{\lambda}^{\dagger}=\frac{1}{i\hbar}\left[ \hat{P}_{\lambda}^{\dagger},\hat{\mathcal{H}} \right]
\end{equation}
are the operators of spin-independent current accumulation at the $\lambda $ orbital \cite{Todorov2002}.

We next introduce the individual components of  $\hat{\mathcal{H}}$ defined in Eq.~(\ref{H_all}) and substitute them into Eqs.~(\ref{spin_torque_operator_1}) and (\ref{spin_current_operator_1}) to disentangle the contributions of $\vec{T}_{\lambda}^{\,\mathrm{S}}$ and $\vec{\Phi}_{\lambda}^{\mathrm{S}}$.
The spin continuity equation of Eq.~(\ref{dsdt_1}) is thus decomposed as
\begin{equation}\label{dsdt_2}
	\frac{d\vec{S}_{\lambda}}{dt} = \vec{T}_{\lambda}^{\,\mathrm{S,XC}} +\vec{T}_{\lambda}^{\,\mathrm{S,SO}}+\vec{\Phi}_{\lambda}^{\mathrm{S,KE}} +\vec{\Phi}_{\lambda}^{\mathrm{S,SO}} ~.
\end{equation}
The first two terms are the exchange spin torque and spin-orbit torque that originate from the exchange (XC) coupling and the spin-orbit (SO) coupling, respectively, i.e.,
\begin{equation}\label{spin_torque_operator_3}
	\vec{T}_{\lambda}^{\,\mathrm{S,XC/SO}} = \frac{1}{2} \left<\left( \hat{P}_{\lambda}^{\dagger}\hatvect{T}^{\mathrm{S,XC/SO}}+\hatvect{T}^{\mathrm{S,XC/SO}}\hat{P}_{\lambda} \right)\right>~
\end{equation}
and
\begin{equation}\label{spin_torque_operator_4}
	\hatvect{T}^{\mathrm{S,XC/SO}}=\frac{1}{i\hbar}\left[ \hatvect{S},\hat{\mathcal{H}}^\mathrm{XC/SO} \right]~.
\end{equation}
Since the kinetic energy and $\hat{\mathcal{V}}_{\lambda}^{0,\mathrm{eff}}$ of Eq.~(\ref{H_all}) are spin-independent, they thus give no contributions to the spin torque.

The last two terms of Eq.~(\ref{dsdt_2}) represent the spin current accumulation attributed by the kinetic energy (KE) and the spin-orbit (SO) coupling, since only these two Hamiltonian components including the momentum operator $\hatvect{p}$ and do not commute with the charge density operator.
In other words, electrons can hop from one orbital to another via the so-called kinetic exchange to contribute a non-zero spin current accumulation including the SOC-induced gradient of Coulomb potential.
This allows us to recast the spin current accumulation in a more intuitive way, namely, the inter-orbital spin current $\vect{J}_{\lambda \lambda'}^{\,\,\mathrm{S}}$ \cite{Todorov2002}:
\begin{align}\label{spin_current_operator_3}
	\vec{\Phi}_{\lambda}^{\mathrm{S,KE/SO}}
	&= -\sum_{\lambda' \ne \lambda}{\vec{J}_{\lambda \lambda'}^{\,\,\mathrm{S,KE/SO}}} \notag
	\\
	&= -\sum_{\lambda' \ne \lambda}{\frac{1}{2} \left<\left( \hatvect{S}\hat{J}_{\lambda \lambda'}^{\,\mathrm{KE/SO}} + \hat{J}_{\lambda \lambda'}^{\dagger\mathrm{KE/SO}}\hatvect{S} \right)\right>}~,
\end{align}
where
\begin{align}\label{spin_current_operator_4}
	\hat{J}_{\lambda \lambda '}^{\,\mathrm{KE/SO}} &\equiv \frac{1}{i\hbar}\left( \hat{P}_{\lambda '}\hat{\mathcal{H}}^{\mathrm{KE/SO}}\hat{P}_{\lambda} -  {P}_{\lambda}\hat{\mathcal{H}}^{\mathrm{KE/SO}}\hat{P}_{\lambda '} \right) \notag
	\\
	\hat{J}_{\lambda \lambda '}^{\dagger \mathrm{KE/SO}} &\equiv \frac{1}{i\hbar}\left( \hat{P}_{\lambda '}^{\dagger}\hat{\mathcal{H}}^{\mathrm{KE/SO}} \hat{P}_{\lambda}^{\dagger} - \hat{P}_{\lambda}^{\dagger}\hat{\mathcal{H}}^{\mathrm{KE/SO}}\hat{P}_{\lambda '}^{\dagger} \right)
\end{align}
are the inter-orbital current operators from $\lambda$ to $\lambda'$ orbitals.
Notice that there exists a minus sign appearing in Eq.~(\ref{spin_current_operator_3}), since the positive value of the spin current ($\vec{J}_{\lambda \lambda'}^{\,\,\mathrm{S}}>0$) \textit{flows-out} from $\lambda$ to $\lambda'$. Therefore, the minus summation ($\vect{\Phi}_{\lambda}^{\mathrm{S}} = -\sum_{\lambda' \ne \lambda}{\vect{J}_{\lambda \lambda'}^{\,\,\mathrm{S}}}$) is an \textit{increase} of ${d\vect{S}_{\lambda}}/{dt}$.

In addition, we have presented the spin current accumulation (or the net flux of spin current) in the discrete summation form. For continuous integral form, we have $\vect{\Phi}^{\mathrm{S}}(\vec{r\,})=-\oint_A{\hat{\mathcal{J}}{}^{\mathrm{S}}\cdot d\vect{A}}$, which accounts the net flux on a close surface $A$. Here $\hat{\mathcal{J}}{}^{\mathrm{S}}$ denotes the spin current tensor in both of the spin and the real space. By employing the divergence theorem, the accumulation in a volume $V$ can also be written as $\vect{\Phi}^{\mathrm{S}}(\vec{r\,})=-\int_V{( \nabla \cdot \hat{\mathcal{J}}{}^{\mathrm{S}} ) dV}$, from which the divergence form $-\nabla \cdot \hat{\mathcal{J}}{}^{\mathrm{S}}$ is more generally used to represent the spin continuity equation \cite{Zhang2004,Ralph2008,Manchon2009,Haney2010,Tsymbal2019,Go2020}.

\subsection{Dynamics of local magnetic moments}
\label{sec:local_moment_dynamics}

Experimentally, the dynamics of magnetic moments (or magnetization) of magnetic materials relates to the measurements of magnetic hysteresis. We can express the dynamics into a classical Landau-Lifshitz-Gilbert (LLG) torque and a quantum-mechanical (QM) torque:
\begin{equation}
	\frac{1}{\gamma}\frac{d\vect{M}_{\lambda}}{dt} = \vec{T}^\mathrm{\,M,LLG}_\lambda + \vec{T}^\mathrm{\,M,QM}_\lambda~.
\end{equation}
Since the time derivative of an angular momentum is defined as torque, i.e., $\vec{\tau}_\lambda = d{\vec{S}^{\,0}_{\lambda}}/dt = -(1/\gamma) d{\vec{M}_\lambda}/dt$ with $\vec{M}_\lambda = -\gamma{\vec{S}^{\,0}_{\lambda}}$, we define a moment torque $\vec{T}^\mathrm{\,M}_\lambda$ such that it is in the same direction of $d\vec{M}_\lambda/dt$.

\begin{figure}
	\centering
	\includegraphics{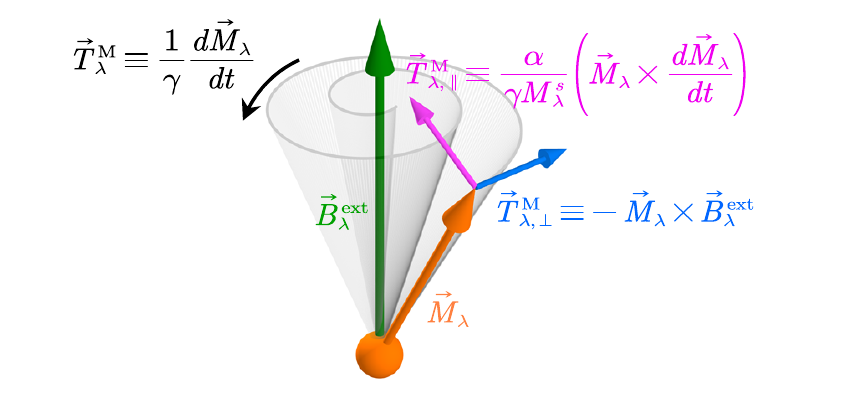}
	\caption{\label{fig_precession}The precession motion of a magnetic moment $\vec{M}_\lambda$ in an external magnetic field $\vec{B}^{\,\mathrm{ext}}_\lambda$ described by Landau-Lifshitz-Gilbert (LLG) equation. Here we define a moment torque $\vec{T}^{\,\mathrm{M}}_\lambda$ whose direction is the same with $d\vec{M}_\lambda/dt$, and $\vec{T}^{\,\mathrm{M}}_{\lambda,\perp}$ along with $\vec{T}^{\,\mathrm{M}}_{\lambda,\parallel}$ are the moment torques in the field-like and the damp-like direction, respectively.}
\end{figure}

The classical LLG torque is expressed by the LLG equation \cite{Landau1992,Gilbert2004}:
\begin{equation}
	\begin{split}
		\vec{T}^\mathrm{\,M,LLG}_\lambda &\equiv \left. \frac{1}{\gamma} \frac{d\vect{M}_{\lambda}}{dt} \right|_{\mathrm{LLG}}
		\\
		&= -\vect{M}_{\lambda}\times \vect{B}_{\lambda}^{\mathrm{ext}}+\frac{\alpha}{\gamma M^s_\lambda}\left( \vect{M}_{\lambda}\times \frac{d\vect{M}_{\lambda}}{dt} \right).
	\end{split}
\end{equation}
The first term of the right-hand side describes the precession motion due to an external magnetic field $\vect{B}_{\lambda}^{\mathrm{ext}}$. The second term with a phenomenological Gilbert damping constant $\alpha$ and a saturation magnetic moment $ M^s_\lambda$ describes the energy loss during the precession motion. The direction of each term is summarized in Fig.~\ref{fig_precession}.

In addition to the classical term, we also require to consider the QM contributions to include the effects of current-induced spin-transfer torque, interlayer exchange coupling, and magnetic anisotropy due to spin-orbit torque (\ref{appendix_tmqm}):
\begin{equation}\label{dmdt_QM_cross_product}
	\vec{T}^\mathrm{\,M,QM}_\lambda \equiv \left. \frac{1}{\gamma} \frac{d\vec{M}_{\lambda}}{dt} \right|_{\mathrm{QM}} =\xi_{\lambda}^{\mathrm{XC}}\vec{M}_{\lambda}\times \vec{S}_{\lambda}~.
\end{equation}
It has been shown that the spin density can be schematically separated into two contributions \cite{Zhang2004}:
\begin{equation}\label{spin_density_2}
	\vec{S}_{\lambda} = \vec{S}^{\,0}_{\lambda} + \delta\vec{S}_{\lambda}~.
\end{equation}
The first term is the local contribution that generates the local spin moment, i.e., $\vec{M}_\lambda = -\gamma{\vec{S}^{\,0}_{\lambda}}$, and the second term is the carriers' contribution to effectively involve the spin-spin and spin-orbit couplings with the environment via exchange interaction. Combining Eqs.~(\ref{dmdt_QM_cross_product}) and (\ref{spin_density_2}), we have
$\vec{T}^\mathrm{\,M,QM}_\lambda = \xi_{\lambda}^{\mathrm{XC}}\vec{M}_{\lambda}\times \delta\vec{S}_{\lambda}$, from which the QM torque can be interpreted as interaction between the local magnetic moment and the carries' spin.

\subsection{Role of spin torque in magnetic dynamics}

In part \ref{sec:spin_continuity} and part \ref{sec:local_moment_dynamics}, we have discussed the dynamics for the system spin density and the local magnetic moment, respectively. To relate exchange spin torque [Eq.~(\ref{spin_torque_operator_3})] to the QM torque [Eq.~(\ref{dmdt_QM_cross_product})], we can expand the commutator and introduce the cross product:
\begin{equation}\label{txc_cross_product}
	\begin{split}
		\vect{T}_{\lambda}^{\,\mathrm{S,XC}}
		&= \langle\hatvect{T}_{\lambda}^{\,\mathrm{S,XC}}\rangle
		\\
		&= \frac{1}{2} \left<\left( \hat{P}_{\lambda}^{\dagger}\hatvect{T}^{\,\mathrm{S,XC}} + \hatvect{T}^{\,\mathrm{S,XC}}\hat{P}_{\lambda} \right)\right>
		\\
		&= \frac{1}{2}\frac{1}{i\hbar} \left<\left( \hat{P}_{\lambda}^{\dagger}\left[ \hatvect{S},\hat{\mathcal{H}}^{\mathrm{XC}} \right] +\left[ \hatvect{S},\hat{\mathcal{H}}^{\mathrm{XC}} \right] \hat{P}_{\lambda} \right)\right>
		\\
		&= -\xi_{\lambda}^{\mathrm{XC}} \frac{1}{2} \left<\left( \hat{P}_{\lambda}^{\dagger}\hatvect{S}\times \vect{M}_{\lambda}+\hatvect{S}\times \vect{M}_{\lambda}\hat{P}_{\lambda} \right)\right>
		\\
		&= -\xi_{\lambda}^{\mathrm{XC}} \left< \frac{1}{2} \left( \hat{P}_{\lambda}^{\dagger}\hatvect{S}+\hatvect{S}\hat{P}_{\lambda} \right)\right> \times \vect{M}_{\lambda}
		\\
		&= -\xi_{\lambda}^{\mathrm{XC}} \vect{S}_{\lambda}\times \vec{M}_{\lambda}~.
	\end{split}
\end{equation}
Comparing with Eq.~(\ref{dmdt_QM_cross_product}), the QM torque can be simplified as the exchange spin torque \cite{Go2020,NL2020}, namely,
\begin{equation}\label{tmqm_equal_tsxc}
	\vec{T}^\mathrm{\,M,QM}_\lambda = \vect{T}_{\lambda}^{\,\mathrm{S,XC}}~.
\end{equation}
In the steady state, $d\vect{S}_{\lambda}(t)/dt=0$ and Eq.~(\ref{dsdt_2}) gives
\begin{equation}\label{Tnet}
	\begin{split}
		\vect{T}_{\lambda}^{\,\mathrm{M,QM}} &= \vect{T}_{\lambda}^{\,\mathrm{S,XC}}
		\\
		&= -\vect{T}_{\lambda}^{\,\mathrm{S,SO}} - \vect{\Phi}_{\lambda}^{\mathrm{S,KE}} - \vect{\Phi}_{\lambda}^{\mathrm{S,SO}}
		\\
		&= -\vect{T}_{\lambda}^{\,\mathrm{S,SO}} + \sum_{\lambda' \ne \lambda}{\left( \vec{J}_{\lambda \lambda'}^{\,\,\mathrm{S,KE}} + \vec{J}_{\lambda \lambda'}^{\,\,\mathrm{S,SO}} \right)} ~.
	\end{split}
\end{equation}
Therefore, in order to account the QM toque for the magnetic moment dynamics, we can compute either $\vect{T}_{\lambda}^{\,\mathrm{S,XC}}$ directly or $\vect{T}_{\lambda}^{\,\mathrm{S,SO}}$, $\vect{\Phi}_{\lambda}^{\mathrm{S,KE}}$ and $\vect{\Phi}_{\lambda}^{\mathrm{S,SO}}$ to see the individual contribution.
Notably, the existence of the intrinsic spin-orbit coupling or non-negligible $\hat{\mathcal{H}}^{\mathrm{SO}}$ not only medias the angular momentum transfer between spin and orbital degree of freedom but also contributes to additional $\vect{T}_{\lambda}^{\,\mathrm{S,SO}}$ and $\vect{\Phi}_{\lambda}^{\mathrm{S,SO}}$ even in the absence of an external electric voltage/current or an applied magnetic field.

In order to study the equality of Eq.~(\ref{tmqm_equal_tsxc}) clearly, we rewrite the local magnetic moment into the form of spin angular momentum, i.e., $\vect{M}_{\lambda} = -\gamma \vect{S}_{\lambda}^\mathrm{\,0}$, such that Eqs.~(\ref{dmdt_QM_cross_product}) and (\ref{txc_cross_product}) using Eq.~(\ref{spin_density_2}) become
\begin{equation}
	\vec{T}^\mathrm{\,S0,QM}_\lambda \equiv \left. \frac{d\vec{S}_{\lambda}^{\,0}}{dt} \right|_{\mathrm{QM}} = \gamma \xi_{\lambda}^{\mathrm{XC}}\vec{S}_{\lambda}^{\,0} \times \delta\vec{S}_{\lambda}
\end{equation}
and
\begin{equation}
	\vect{T}_{\lambda}^{\,\mathrm{S,XC}} = \gamma \xi_{\lambda}^{\mathrm{XC}} \delta\vect{S}_{\lambda}\times \vec{S}_{\lambda}^{\,0}~.
\end{equation}
We can find that
\begin{equation}
	\vec{T}^\mathrm{\,S0,QM}_\lambda = -\vect{T}_{\lambda}^{\,\mathrm{S,XC}}~,
\end{equation}
which can be interpreted that the increase (decrease) of the local spin $\vect{S}_{\lambda}^\mathrm{\,0}$ is equal to the decrease (increase) of the carriers' spin $\delta\vect{S}_{\lambda}$. Such relation provides a basic idea of spin-transfer effect between the local spin and the carriers' spin via the exchange interaction.

\begin{figure}
	\centering
	\includegraphics{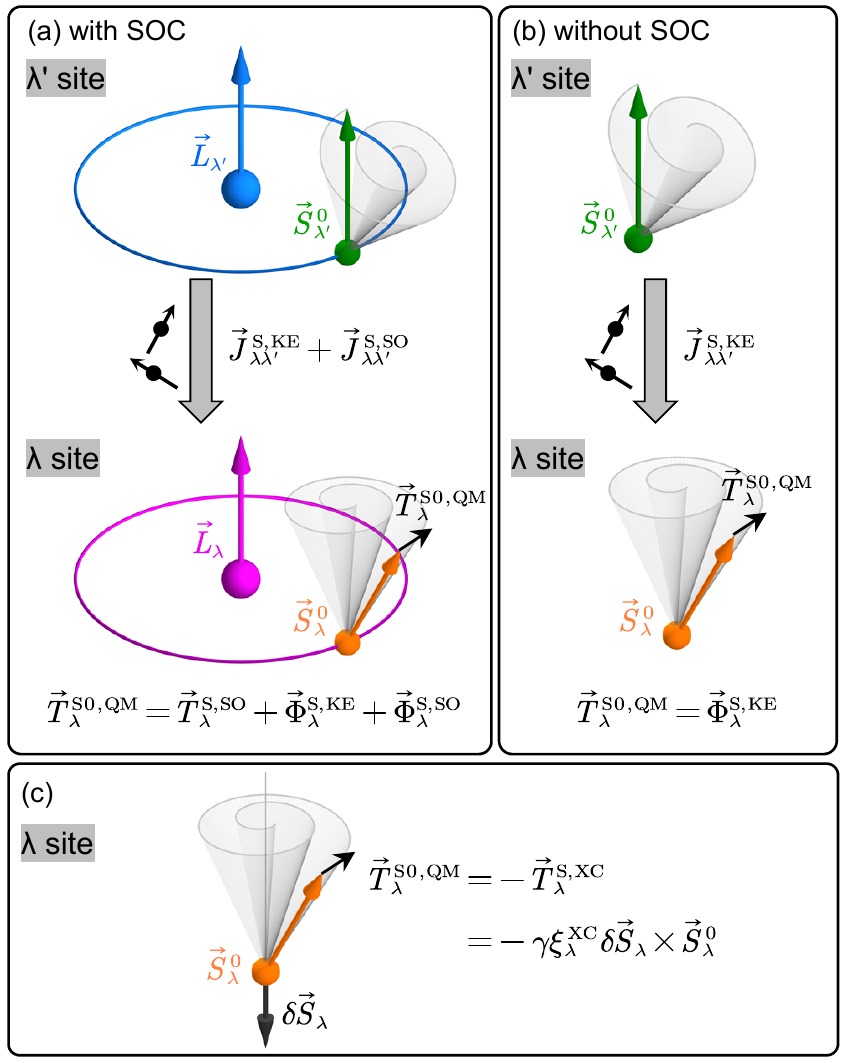}
	\caption{\label{fig_relation}A schematic plot for the relation between the local spin $\vec{S}_{\lambda}^{\,0}$, the environmental spins $\vec{S}_{\lambda'}^{\,0}$, the environmental orbitals $\vec{L}_{\lambda'}$, and the local orbital $\vec{L}_{\lambda}$. (a) The interactions with the environmental spins and orbitals are driven by the kinetic part and spin-orbit part of the spin current, i.e., $\vec{J}^{\,\,\mathrm{S,KE}}_{\lambda\lambda'}$ and $\vec{J}^{\,\,\mathrm{S,SO}}_{\lambda\lambda'}$, respectively. The magnetic anisotropy effect generates a spin-orbit torque $\vect{T}_{\lambda}^{\,\mathrm{S,SO}}$ exerting on the local spin. (b) In the absence of SOC, only the kinetic part contributes the spin torque. (c) The whole effects on the local spin can be views as an effective carriers' spin $\delta\vec{S}_\lambda$ that generates a exchange spin torque $\vect{T}_{\lambda}^{\,\mathrm{S,XC}}$ on $\vec{S}_{\lambda}^{\,0}$.}
\end{figure}

In the steady state, which demands the spin conservation of the system, Eq.~(\ref{Tnet}) also reveals the relation between the spin current accumulation ($\vect{\Phi}_{\lambda}^{\mathrm{S,KE}}$ and $\vect{\Phi}_{\lambda}^{\mathrm{S,SO}}$) and the angular momentum variation ($\vect{T}_{\lambda}^{\,\mathrm{S,XC}}$ and $\vect{T}_{\lambda}^{\,\mathrm{S,SO}}$) as summarized in Fig.~\ref{fig_relation}.
The interaction between the environmental spins ($\vec{S}_{\lambda'}^{\,0}$) and the local spin ($\vec{S}_{\lambda}^{\,0}$) can be viewed as a spin current $\vec{J}^{\,\,\mathrm{S,KE}}_{\lambda\lambda'}$ between them, which is driven by the electrons with nonzero kinetic energy.
In the meantime, the moving electrons also suffer a SOC generated by the environmental orbitals ($\vec{L}_{\lambda'}$), which appears as an effective magnetic field, and generates an effective SO contributed spin current $\vec{J}^{\,\,\mathrm{S,SO}}_{\lambda\lambda'}$.
In the consideration of SOC that attributes the magnetic anisotropy, if the direction of the local spin is not in the magnetic easy-axis, then it experiences a spin-orbit torque $\vect{T}_{\lambda}^{\,\mathrm{S,SO}}$ from the local orbital ($\vec{L}_{\lambda}$) to flip back to the easy-axis.
The overall effects on the local spin $\vec{S}^{\,0}_\lambda$ can be viewed as an effective carriers' spin $\delta \vec{S}_\lambda$ that is not parallel to the local spin, and exerts a net torque $\vect{T}_{\lambda}^{\,\mathrm{S,XC}}$ on it by the exchange interaction.

\section{First-principles implementation}
\label{sec:implementation}

\subsection{Localized orbital representation}

We next construct the matrix representations of the spin torque and spin current accumulation by linear combination of atomic orbitals (LCAO) approach. The expectation value of the spin current accumulations and the spin torques will be derived. Note that, in this study, an element of a matrix $\mat{O}$ is defined as $\mat{O}_{mn}\equiv ( \mat{O} ) _{mn}$, and a one of the operated matrix is $\mat{O}_{mn}^{X}\equiv ( \mat{O}^X ) _{mn}$ with $X=\left\{ -1,\dagger ,\mathrm{T} \right\}$ being the inverse, Hermitian, or transpose operation. For a spin-polarized system, a matrix element $\mat{O}_{\mu \nu}$ with the orbital indices $\mu$ and $\nu$ is a $2\times 2$ submatrix in spin space. The system is expanded by a finite set of nonorthogonal atomic orbitals $\left\{ |\mu \rangle \right\} $ with the overlapping matrix element as $\mat{\Omega }_{\mu \nu}=\langle \mu |\nu \rangle $ and the Hamiltonian matrix element as $\mat{H}_{\mu \nu}=\langle \mu |\hat{\mathcal{H}}|\nu \rangle $. The individual contributions $\mat{H}_{\mu \nu}^0$, $\mat{H}_{\mu \nu}^{\mathrm{XC}}$ and $\mat{H}_{\mu \nu}^{\mathrm{SO}}$ are obtained as described in \hyperref[appendix_h_components]{\ref*{appendix_h_components}}. Using the overlapping matrix elements, the identity operator can be expanded as
\begin{equation}
	\hat{I}=\sum_{\mu \nu}{|\mu \rangle \mat{\Omega }_{\mu \nu}^{-1}\langle \nu |}.
\end{equation}
Due to the nonorthogonal property, the projection operators to the $\lambda$ orbital have two types of form \cite{Soriano2014}:
\begin{align}
	\hat{P}_{\lambda} &\equiv \sum_{\nu}{|\lambda \rangle \mat{\Omega }_{\lambda \nu}^{-1}\langle \nu |} \\ \hat{P}_{\lambda}^{\dagger} &\equiv \sum_{\nu}{|\nu \rangle \mat{\Omega }_{\nu \lambda}^{-1}\langle \lambda |}~.
\end{align}
These operators in terms of matrix representation can be used to rewrite the formulas.

We also need a density matrix to compute the expectation values. The eigenvalue method is first presented here and the Green’s function approach will be discussed in the next section. Using the basis set, the eigenket and eigenbra of a state $n$ are respectively expanded as
\begin{align}
	|\psi _n\rangle &= \sum_{\nu}{|\nu \rangle \mat{C}_{\nu n}} \\
	\langle \psi _n| &= \sum_{\mu}{\mat{C}_{n\mu}^{\dagger}\langle \mu |}~,
\end{align}
where $\mat{C}_{\nu n}$ and $\mat{C}_{n\mu}^{\dagger}$ are the expansion coefficients. In steady state, the coefficients can be solved by the eigenvalue equations:
\begin{equation}
	\begin{gathered}
		\mat{HC}=\mat{\Omega C\Lambda }\\\mat{C}^{\dagger}\mat{H}=\mat{\Lambda C}^{\dagger}\mat{\Omega }\\\mat{\Lambda }=\mathrm{diag}\left( \varepsilon _1,\varepsilon _2,\cdots ,\varepsilon _n \right),
	\end{gathered}
\end{equation}
where $\mat{\Lambda }$ is a diagonal matrix filled with the eigenenergy $\varepsilon _n$. The elements of the density matrix $\mat{D}_{\mu \nu}$ and the energy density matrix $\mat{D}_{\mu \nu}^{\mathrm{E}}$ are computed by
\begin{align}
	\mat{D}_{\mu \nu} &= \sum_n{f_n\mat{C}_{\mu n}\mat{C}_{n\nu}^{\dagger}} \\ \mat{D}_{\mu \nu}^{\mathrm{E}} &= \sum_n{f_n\varepsilon _n\mat{C}_{\mu n}\mat{C}_{n\nu}^{\dagger}},
\end{align}
where $f_n$ is an occupation factor. The matrices $\mat{H}$, $\mat{\Omega}$ and $\mat{D}$ required to compute the spin torque and spin current accumulation can be constructed, for example, by first-principles calculations \cite{Taylor2001,Haney2007} or by tight-binding models \cite{Theodonis2006,Tang2010}.

Next, we show the expectation values of the spin torque and spin current accumulation by using the LCAO represented operators above. We first have the matrix representation of the spin angular momentum operator as
\begin{equation}
	{\vectmat{S}}_{\mu \nu}=\langle \mu |\hatvect{S}|\nu \rangle = \frac{\hbar}{2}\vectmat{\sigma}\mat{\Omega}_{\mu \nu},
\end{equation}
where $\vectmat{\sigma}=\left( \mat{\sigma }^x,\mat{\sigma }^y,\mat{\sigma }^z \right) $ is the vector of $2 \times 2$ Pauli matrices. For the spin torque operator, the matrix elements are computed by
\begin{equation}\label{spin_torque_element}
	\vectmat{T}{}_{\mu \nu}^{\mathrm{S}}=\langle \mu |\hatvect{T}{}^{\mathrm{S}}|\nu \rangle =\frac{1}{i\hbar}\left[ \frac{\hbar}{2}\vectmat{\sigma},\mat{H}_{\mu \nu} \right],
\end{equation}
and the expectation value of the $\lambda$ orbital is derived as
\begin{equation}
	\begin{split}\label{spin_torque_expectation}
		\vect{T}_{\lambda}^{\mathrm{S}} &= \langle \hatvect{T}_{\lambda}^{\mathrm{S}} \rangle \\&=\frac{1}{2}\left( \langle \hat{P}_{\lambda}^{\dagger}\hatvect{T}^{\mathrm{S}} \rangle +\langle \hatvect{T}^{\mathrm{S}}\hat{P}_{\lambda}\rangle \right) \\ &= \frac{1}{2}\mathrm{Tr}_{\sigma}\left( \vectmat{T}{}^{\mathrm{S}}\mat{D}+\mat{D}\vectmat{T}{}^{\mathrm{S}} \right) _{\lambda \lambda}.
	\end{split}
\end{equation}
For the spin current, we first define two quantities
\begin{align}
	\vectmat{F}{}_{\lambda \lambda'}^{(1)} &\equiv \frac{1}{i\hbar}\frac{\hbar}{2}\left[ \mat{D}_{\lambda \lambda'}^{(2)}\vectmat{\sigma}\mat{H}_{\lambda' \lambda}^{(1)} + \mat{D}_{\lambda \lambda'}^{(1)}\mat{H}_{\lambda' \lambda}^{(2)}\vectmat{\sigma} \right]
	\\
	\vectmat{F}{}_{\lambda \lambda'}^{(2)} &\equiv \frac{1}{i\hbar}\frac{\hbar}{2}\left[ \vectmat{\sigma}\mat{H}_{\lambda \lambda'}^{(1)}\mat{D}_{\lambda' \lambda}^{(2)} + \mat{H}_{\lambda \lambda'}^{(2)}\vectmat{\sigma}\mat{D}_{\lambda' \lambda}^{(1)} \right],
\end{align}
where we have used the definition
\begin{equation}
	\begin{array}{c}
		\mat{H}^{\left( 1 \right)}\equiv \mat{\Omega }^{-1}\mat{H}\\
		\mat{H}^{\left( 2 \right)}\equiv \mat{H\Omega }^{-1}\\
	\end{array}\quad\quad
	\begin{array}{c}
		\mat{D}^{\left( 1 \right)}\equiv \mat{\Omega D}\\
		\mat{D}^{\left( 2 \right)}\equiv \mat{D\Omega }\\
	\end{array},
\end{equation}
and the expectation value is derived as
\begin{equation}\label{spin_current_3}
	\begin{split}
		\vect{J}{}_{\lambda \lambda'}^{\,\,\mathrm{S}} &= \langle \hatvect{J}{}_{\lambda \lambda'}^{\,\,\mathrm{S}} \rangle \\ &= \frac{1}{2}\left( \langle \hatvect{S}\hat{J}_{\lambda \lambda'} \rangle + \langle \hat{J}_{\lambda \lambda'}^{\dagger}\hatvect{S} \rangle \right) \\ &= \frac{1}{2}\mathrm{Tr}_{\sigma}\left( \vectmat{F}{}_{\lambda \lambda'}^{\left( 1 \right)}-\vectmat{F}{}_{\lambda \lambda'}^{\left( 2 \right)} \right).
	\end{split}
\end{equation}
Finally, the spin current accumulation is
\begin{equation}
	\vect{\Phi}_{\lambda}^{\mathrm{S}} = \langle \hatvect{\Phi}_{\lambda}^{\mathrm{S}} \rangle = -\sum_{\lambda' \ne \lambda}{\langle \hatvect{J}_{\lambda \lambda'}^{\,\mathrm{S}} \rangle}.
\end{equation}

Here the spin current equations seem not intuitive because of the existing $\mat{\Omega}$ and $\mat{\Omega}^{-1}$. If the system is expanded by an orthogonal basis set, i.e., $\mat{\Omega} = \mat{\Omega}^{-1} = \mat{I}$, then Eq.~(\ref{spin_current_3}) is simplified as
\begin{equation}\label{spin_current_orth}
	\vect{J}{}_{\lambda \lambda'}^{\,\,\mathrm{S,orth}} = \frac{1}{2}\mathrm{Tr}_{\sigma}\left( \mat{D}_{\lambda \lambda'}\vectmat{J}{}_{\lambda' \lambda}^{\mathrm{S,orth}} - \vectmat{J}{}_{\lambda \lambda'}^{\mathrm{S,orth}}\mat{D}_{\lambda' \lambda} \right),
\end{equation}
where
\begin{equation}\label{spin_current_orth2}
	\vectmat{J}{}_{\lambda' \lambda}^{\mathrm{S,orth}} = \frac{1}{i\hbar} \left\lbrace \frac{\hbar}{2} \vectmat{\sigma}, \mat{H}_{\lambda' \lambda} \right\rbrace.
\end{equation}
It shows that the commutator in the calculation of spin torque [Eq.~(\ref{spin_torque_element})] is replaced by the anti-commutator in the spin current calculation [Eq.~\ref{spin_current_orth2}]. Also, the addition property of Eq.~(\ref{spin_torque_expectation}) becomes the subtraction in Eq.~(\ref{spin_current_orth}). Such relation can also be found in the representation of wave function formalism \cite{Go2020}.

\subsection{NEGF formalism}

So far, the spin torque and spin current accumulation have been derived in terms of the density matrix. For applications using two-probe structures, we use the Keldysh nonequilibrium Green’s function (NEGF) formalism to compute the density matrix \cite{Datta1995,Taylor2001,Brandbyge2002}. In steady state, the retarded Green’s function is computed via
\begin{equation}
	\mat{G}^{r}\left( E \right) =\left[ E \mat{\Omega }-\mat{H}-\mat{\Sigma }_{\mathrm{B}}^{r}\left( E^+ \right) -\mat{\Sigma }_{\mathrm{T}}^{r}\left( E^+ \right) \right] ^{-1},
\end{equation}
where $E^+=E+i\eta$ with $\eta$ being an infinitesimal positive number. Also, $\mat{G}^{a}=\left( \mat{G}^{r} \right) ^{\dagger}$ is the advanced Green’s function and $\mat{\Sigma }_{\mathrm{B(T)}}^{r}$ is the self-energy matrix of the bottom (top) electrode. Given the Keldysh Green’s function $\mat{G}^<=\mat{G}^r\mat{\Sigma }^<\mat{G}^a$ and the lesser self-energy $\mat{\Sigma }^<=i \left( f_\textrm{B}\mat{\Gamma }_\textrm{B} + f_\textrm{T}\mat{\Gamma }_\textrm{T} \right)$ with $\mat{\Gamma }_{\textrm{B(T)}}=i ( \mat{\Sigma }_{\textrm{B(T)}}^{r}-\mat{\Sigma }_{\textrm{B(T)}}^{a} )$  being the broadening function of the bottom (top) electrode, we have the density matrix and the energy density matrix as
\begin{align}
	\mat{D}_{\mu \nu} &= -i \int{\frac{dE}{2\pi} \mat{G}_{\mu \nu}^{<}\left( E \right)} \\ \mat{D}_{\mu \nu}^{\mathrm{E}} &= -i \int{\frac{dE}{2\pi}E \mat{G}_{\mu \nu}^{<}\left( E \right)}~.
\end{align}
If the system is in equilibrium, the Keldysh Green’s function can be simply expressed as \cite{Datta1995}
\begin{equation}
	\mat{G}^{<,\mathrm{eq}}\left( E \right) =f\left( E \right) \left[ \mat{G}^a\left( E \right) -\mat{G}^r\left( E \right) \right]
\end{equation}
and the density matrices are thus
\begin{align}
	\mat{D}_{\mu \nu}^{\mathrm{eq}}&=-i\int{\frac{dE}{2\pi} f\left( E \right) \left[ \mat{G}_{\mu \nu}^{a}\left( E \right) -\mat{G}_{\mu \nu}^{r}\left( E \right) \right]}\\\mat{D}_{\mu \nu}^{\mathrm{E,eq}}&=-i\int{\frac{dE}{2\pi}E f\left( E \right) \left[ \mat{G}_{\mu \nu}^{a}\left( E \right) -\mat{G}_{\mu \nu}^{r}\left( E \right) \right]}
\end{align}
where $f\left( E \right)$ is the Fermi-Dirac distribution function. The energy integration of the equilibrium density matrices can be performed efficiently by integrating along a complex contour \cite{Taylor2001}.

\section{Calculation details}

\begin{figure}
	\centering
	\includegraphics{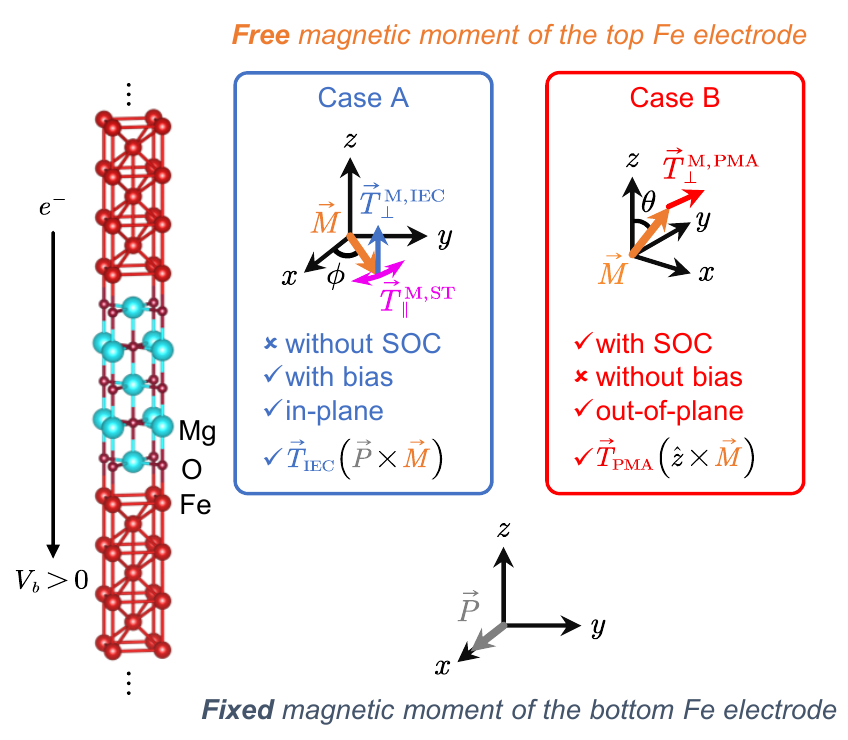}
	\caption{\label{fig1}The atomic structure of the Fe/MgO/Fe MTJ. The magnetic moments of the bottom Fe electrode are fixed and the top are freely rotated either in $x$-$y$ plane (in-plane) or $z$-$x$ plane (out-of-plane). In case A, the system is carried out without SOC but under an external bias to study the current-driven spin-transfer torque. In case B, the SOC is considered but without an external bias to study the interfacial magnetic anisotropy by spin-orbit torque.}
\end{figure}

A conventional magnetic tunnel junction Fe/MgO/Fe is employed to study the spin torque and spin current accumulation, and the structure is the same with our previous calculation \cite{Huang2021}. We construct the system by a two-probe structure, in which the MgO insulator is sandwiched by two semi-infinite bcc-Fe electrodes as shown in Fig.~\ref{fig1}. The composition Fe(7)/MgO(5)/Fe(6), where the numbers are the number of atomic layers, is chosen as the central scattering region. The structure of the electrodes and the scattering region are optimized using the DFT based VASP package \cite{Kresse1996,Kresse1999} with the projector augmented wave (PAW) pseudopotential and PBE exchange correlation functional \cite{Perdew1996}. The energy cutoff for the plane-wave-basis set is 700 eV, and the Brillouin zone sampling is $31\times 31\times 1$ Monkhorst-Pack grid. The lattice constant of the transverse $x$-$y$ plane is fixed with the optimized Fe bulk value, i.e., $a=b=2.84$ {\AA}, and the optimized length of the longitudinal $z$ direction is $c=30.32$ {\AA}. Also, the optimized distance between the interfacial Fe and MgO atomic layers is $2.24$ {\AA}.

Given the optimized structure, we next obtain the self-consistent electronic structure by using the DFT-NEGF based Nanodcal package \cite{Taylor2001,Waldron2007}, where the double-$\zeta$ double-polarized LCAO basis set is used to construct the Hamiltonian and overlapping matrices. Meanwhile, the exchange-correlation functional is based on the local spin density approximation (LSDA), in which the direction of a local magnetic moment is locally related to the spin-splitting \cite{Haney2008,Ralph2008}. The Brillouin zone sampling of the semi-infinite electrodes and the scattering region are respectively taken as $15\times 15\times 100$ and $15\times 15\times 1$ Monkhorst-Pack grid.

In this study, we consider two cases in the non-collinear magnetic configurations as presented in Fig.~\ref{fig1}, where the direction of magnetic moment of the bottom Fe electrode, $\vec{P}$, is fixed along the $x$-axis, and the top, $\vec{M}$, is freely rotated either in the $z$-$x$ plane (out-of-plane) by an angle $\theta$ or in the $x$-$y$ plane (in-plane) by an angle $\phi$. For case A, the spin torque calculation is carried out without SOC but under an external bias, in which the electrons flow from the top to bottom when $V_b>0$, with the in-plane non-collinear magnetic configurations. As for case B, the SOC is considered for out-of-plane non-collinear magnetic configurations at equilibrium ($V_b=0$). The accuracy of the self-consistent Hamiltonian and the band energy are chosen as $10^{-6}$ eV for case A but reduced to $10^{-4}$ eV for case B, since the latter one is more difficult to achieve self-consistent due to the intense charge transfer between Fe-$d_{z^2}$ and O-$p_z$ orbitals at the interface.

The physical quantities including the spin current accumulations and the spin torques are calculated by our developed \textsc{JunPy} package using the self-consistent Hamiltonian above \cite{Tang2018}. Also, the divide-and-conquer technique is used to study the atom-resolved quantities in the semi-infinite electrodes \cite{Huang2021}. The complex contour energy integral method is employed for the equilibrium part of the density matrix, and the nonequilibrium part is taken over the real axis \cite{Taylor2001}. For the Brillouin zone sampling with $100 \times 100 \times 1$ mesh grid, the crystal symmetry for the cases without SOC is used to reduce the sampling points while those with SOC are not reduced because the physical values in the k-space are not assumed to be symmetric \cite{Lee2015}.

\section{Results and discussion}

To study the net effects of the spin torque and spin current accumulation within a finite thickness, the cumulative spin torques and cumulative spin current accumulations are obtained by summing the contributions of all atomic orbitals $\lambda$ within a thickness $t_m=2$ nm of the top Fe electrode, i.e.,
\begin{equation}
	\begin{split}
		\vec{T}^{\,\mathrm{S,XC/SO}} &= \sum_{\lambda \in t_m}\vec{T}_{\lambda}^{\,\mathrm{S,XC/SO}}
		\\
		\vec{\Phi}^{\mathrm{S,KE/SO}} &= \sum_{\lambda \in t_m}\vec{\Phi}_{\lambda}^{\mathrm{S,KE/SO}}~.
	\end{split}
\end{equation}
Also, the torques and the accumulations can usually be decomposed into the damp-like ($\parallel$) and the field-like ($\perp$) directions as described in Fig.~\ref{fig_precession}.

\subsection{Current-driven STT without SOC}

\begin{figure}
	\centering
	\includegraphics{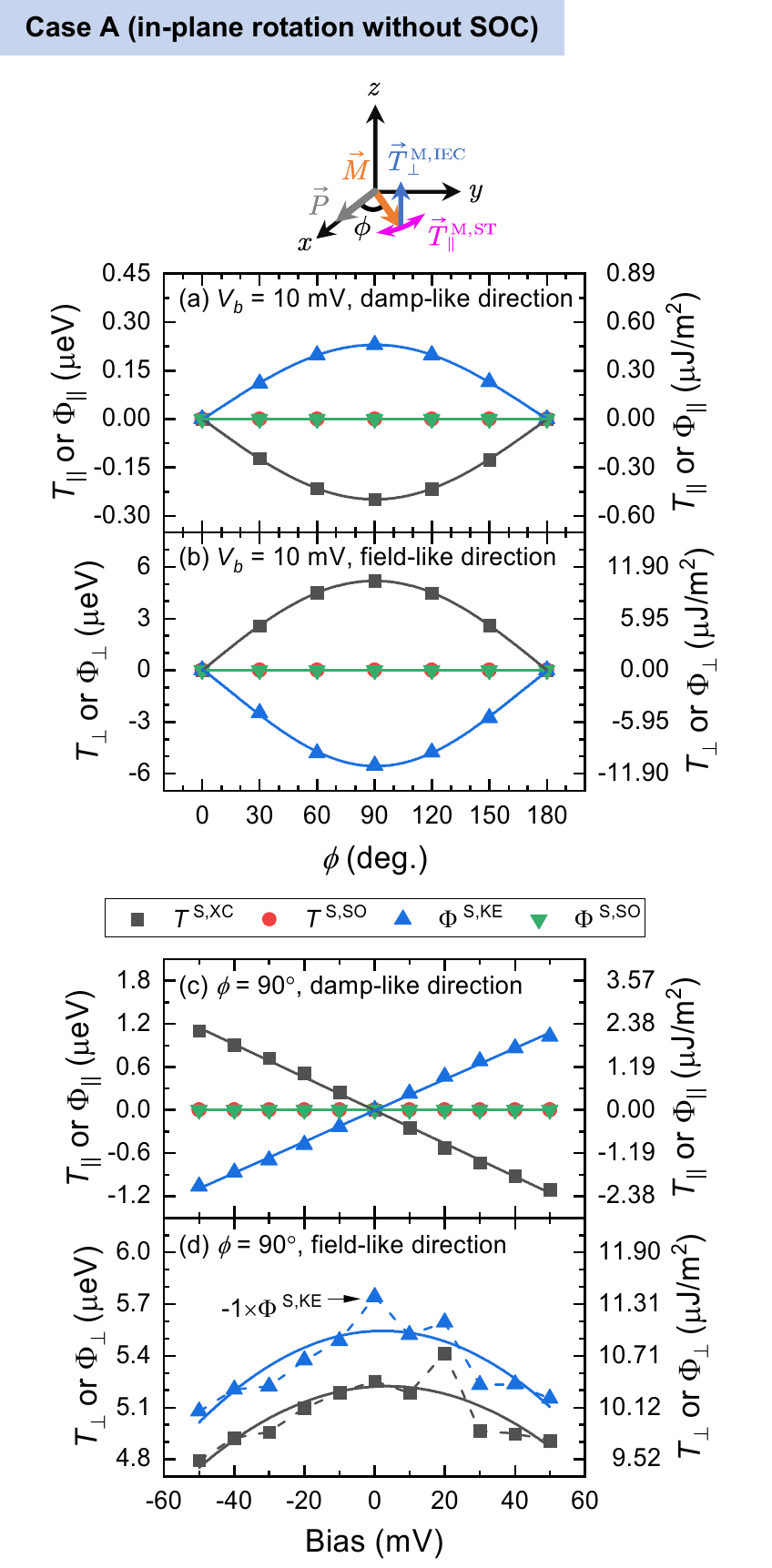}
	\caption{\label{fig_caseA}The in-plane angle $\phi$ dependence of the total spin torques $T^{\mathrm{S,XC/SO}}$ and the total spin current accumulations $\Phi^{\mathrm{S,KE/SO}}$, which are summed over the contributions of all atomic orbital $\lambda$ within the thickness $t_m$ = 2 nm of the top Fe electrode, in (a) the damp-like and (b) the field-like directions for case A without SOC but under an external bias of $V_b=10$ mV. The solid lines are fitted by a sine function. The bias dependence in (c) the damp-like and (d) the field-like directions for case A with an in-plane rotational angle $\phi=90^\circ$. Here the positive magnitudes of $T_{\|}$ and $T_{\bot}$ are in the damp-like ($\vec{M} \times \vec{P} \times \vec{M}$) and the field-like ($\vec{P} \times \vec{M}$) directions, respectively. The lines of the damp-like contributions are fitted by a linear function, and the field-like is fitted by a quadratic function.}
\end{figure}

We first testify the spin torque and spin current accumulation calculations for case A shown in Fig.~\ref{fig1}, where the magnetic moments of the bottom Fe electrode $\vec{P}$ is fixed to $x$-axis and the top $\vec{M}$ is free to rotate by an angle $\phi$ in the $x$-$y$ plane (in-plane).
In the absence of SOC for Eq.~(\ref{Tnet}), the exchange spin torque should be equal to the kinetic spin current accumulation:
\begin{equation}\label{T_caseA}
	\vect{T}^{\,\mathrm{M,QM}} = \vect{T}^{\,\mathrm{S,XC}} = -\vect{\Phi}^{\mathrm{S,KE}}~,
\end{equation}
which can be decomposed  into the so-called current-driven spin-transfer torque (STT, $T_\parallel$) and the field-like torque (FLT, $T_\perp$) acting on the top magnetic moment $\vec{M}$ to align them toward the damping ($\vec{M} \times \vec{P} \times \vec{M}$) and the precession ($\vec{P} \times \vec{M}$) direction, respectively.

Case A is the same with previous theoretical models for conventional MgO-based MTJ calculations \cite{Theodonis2006,Stile2008,Tang2010,Guo2011} and is comparable with experimental measurements of real MTJs with in-plane magnetization \cite{Kubota2008,Sankey2008}.
We present in Figs.~\ref{fig_caseA} (a) and (b) the in-plane angle $\phi$ dependence of the cumulative spin torque $T^{\,\mathrm{S,XC/SO}}$, and the cumulative spin current accumulation $\Phi^{\mathrm{S,KE/SO}}$ within a thickness $t_m = 2$ nm of the top Fe electrode under an external bias of $V_b=10$ mV.
This is because our previous \textsc{JunPy}+DC calculation has demonstrated that the cumulative spin torques reach maximum near the MgO/Fe interface and suggest that the Fe layer thinner than 2 nm is better to preserve the current-driven magnetization switching process \cite{Huang2021}.

Equation~(\ref{T_caseA}) can be easily verified from Figs.~\ref{fig_caseA} (a) and (b). Namely, the negative magnitude of the kinetic spin current accumulation is the main contribution of the exchange spin torque, since the other two SOC-induced terms are zero in the absence of SOC. These results agree with the general understanding in the spin-transfer effect of insulator-based MTJs \cite{Stiles2002,Slonczewski2005,Ralph2008}, which states that in nonzero bias the spin current flowing from the fixed layer may transfer the spin angular momentum to the free layer and then results in the spin accumulation at the interface to switch the magnetization direction of the free layer.

To identify the computational accuracy, we display in Figs.~\ref{fig_caseA} (c) and (d) the bias dependence of the cumulative $T^{\mathrm{S,XC}}$ and the cumulative $ \Phi^{\mathrm{S,KE}} $ within a thickness $t_m=2$ nm of the top Fe electrode with an in-plane rotational angle $\phi=90^\circ$.
For an external bias of $V_b=20$ mV, the calculated $T^{\mathrm{S,XC}}_{\parallel} = -1.03~\mu\mathrm{J/m}^2$ and fitted $T^{\mathrm{S,XC}}_{\perp}(20~\mathrm{mV}) - T^{\mathrm{S,XC}}_{\perp}(0~\mathrm{mV}) = -0.08~\mu\mathrm{J/m}^2$ agree with Kubota's experimental values of $T_{\parallel} \sim 3.9~\mu\mathrm{J/m}^2$ and $T_{\perp} \sim -0.7~\mu\mathrm{J/m}^2$ \cite{Kubota2008}, since the thickness of the MgO barrier around 1.1 nm used in this study is similar to the experimental setup of 1.2 nm.
Here the minus sign of experimental $T_{\|}$ is attributed from Kubota's definition of $\vec{S}_2 \times \vec{S}_1 \times \vec{S}_2$, where $\vec{S}_{1(2)}$ denotes the spin angular momentum of the fixed (free) Fe electrode and exhibits an opposite sign conversion of $\vec{M} = -\gamma\vec{S}$ \cite{Kubota2008}.
This quantitative agreement between the theory and the experiment along with the linear and quasi-quadratic bias dependence of $T_{\|}$ and $T_{\bot}$ suggest the possibility of predictive calculations for other magnetic heterojunctions.

We next discuss case A at equilibrium ($V_b=0$). The non-zero $T^{\mathrm{S,XC}}_{\perp}(0~\mathrm{mV})=10.5$ $\mu$J/m$^2$ originated from the interlayer exchange coupling (IEC) \cite{Tang2009,Tang2010}, i.e., the exchange interaction between the two Fe electrodes, can be related to the energy in the form of \cite{Slonczewski1993,Bruno1995}
\begin{equation}\label{eq_u_iec}
	E^{\mathrm{IEC}}(\phi) = -K_1^{\mathrm{IEC}} \cos\phi -K_2^{\mathrm{IEC}} \cos^2\phi +\cdots~,
\end{equation}
where $K_1^{\mathrm{IEC}}$ and $K_2^{\mathrm{IEC}}$ are the bilinear and biquadratic coupling constants, respectively.
The corresponding field-like torque can be obtained by the energy differentiation as
\begin{equation}\label{eq_t_iec}
	\begin{split}
		T^{\mathrm{IEC}}_{\perp} \left( \phi \right) &\equiv - \frac{\partial E^{\mathrm{IEC}}}{\partial \phi}
		\\
		&= K_{1}^{\mathrm{IEC}}\sin\phi + K_2^{\mathrm{IEC}}\sin2\phi +\cdots~,
	\end{split}
\end{equation}
As presented in Fig.~\ref{fig_caseA} (b), the dominated $\sin\phi$ dependency indicates the fact of $K_{1}^{\mathrm{IEC}} \gg K_{2}^{\mathrm{IEC}}$, which is observed in the previous theoretical and experimental results of magnetic tunnel junctions \cite{Tang2010,Kubota2008}.

\subsection{Interfacial SOT and PMA at equilibrium}

\begin{figure}
	\centering
	\includegraphics{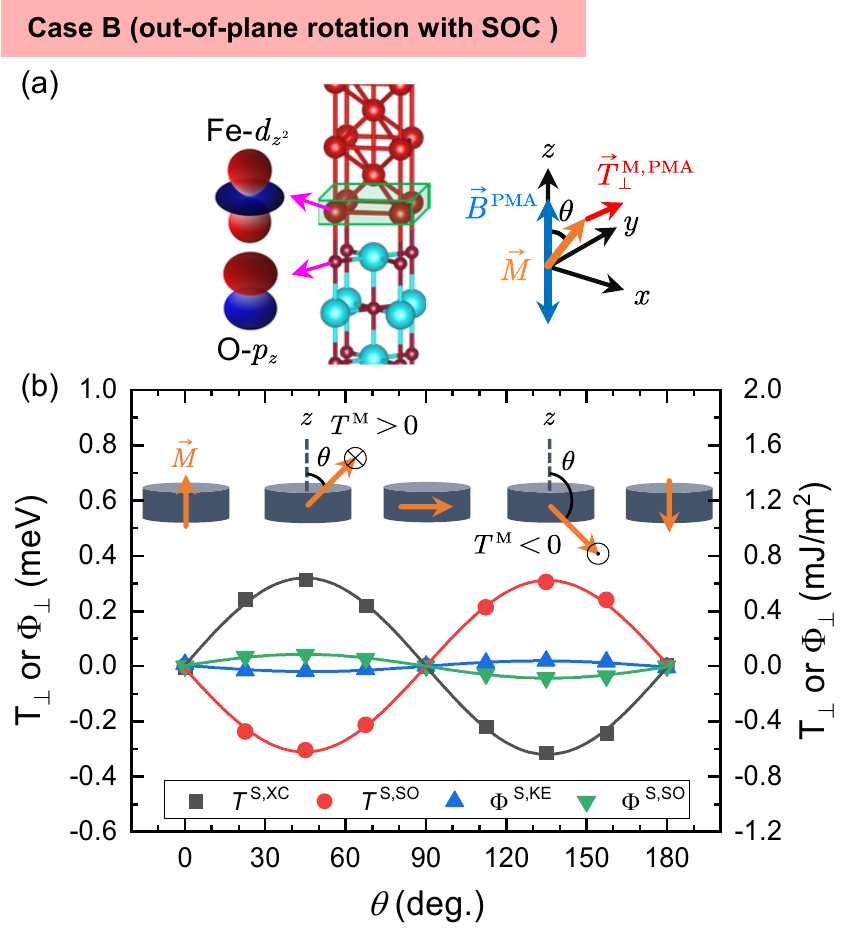}
	\caption{\label{fig_caseC}(a) Side view of case B with SOC at equilibrium ($V_b=0$). The $\vec{B}^\mathrm{PMA}$ field denotes the out-of-plane uniaxial magnetic anisotropy field. (b) The out-of-plane angle $\theta$ dependence of the total spin torques $T^{\mathrm{S,XC/SO}}$ and the total spin current accumulations $\Phi_{\lambda}^{\mathrm{S,KE/SO}}$ in the field-like direction, which are the sum over the contributions of all atomic orbital $\lambda$ within the thickness $t_m=2$ nm of the top Fe electrode. All of the four solid lines are fitted by a sine function. Here the positive magnitudes of $T_{\perp}$ and $\Phi_{\perp}$ are in the $y$ direction corresponding to the precession direction ($-\vec{M} \times \vec{B}^\mathrm{PMA}$).}
\end{figure}

In the consideration of SOC at equilibrium ($V_b=0$), the out-of-plane rotated $\vec{M}$ of case B by an angle $\theta$ is proposed in Figs.~\ref{fig_caseC} (a).
The interfacial PMA is studied, not choosing energy parameters but deriving accurate DFT-based SOT calculation.
The angular dependence of the cumulative spin torque $T^{\,\mathrm{S,XC/SO}}_\perp$ and the cumulative spin current accumulation $\Phi^{\mathrm{S,KE/SO}}_\perp$ within the thickness $t_m = 2$ nm of the top Fe electrode in the field-like ($\perp$) direction are presented in Figs.~\ref{fig_caseC} (b).
Note that there are only nonzero field-like contributions at equilibrium.

From former study \cite{YangPRB2011}, it has been known that the strong SOC between Fe-$d_{z^2}$ and O-$p_z$ orbitals attributes to an effective magnetic field, $\vect{B}^\mathrm{PMA}$, in the uniaxial directions (i.e., $\pm z$ axes) as presented in Fig.~\ref{fig_caseC} (a).
The field exerts a field-like moment torque, $\vect{T}^{\,\mathrm{M,PMA}} = -(\vect{M} \times \vect{B}^\mathrm{PMA})$, on $\vect{M}$ of the top Fe electrode.
The potential energy associated with the system can be described in a general form of the uniaxial magnetic anisotropy as
\begin{equation}\label{eq_u_pma}
	E^{\mathrm{PMA}}\left( \theta \right) = E_{0}^{\mathrm{PMA}} + K_{2}^{\mathrm{PMA}}\sin^2\theta +\cdots~,
\end{equation}
and the corresponding magnitude of field-like moment torque is
\begin{equation}\label{eq_t_pma}
	T^{\mathrm{M,PMA}}_\perp \left( \theta \right) \equiv \frac{\partial E^{\mathrm{PMA}}}{\partial \theta}=K_{2}^{\mathrm{PMA}}\sin 2\theta +\cdots~.
\end{equation}
In the consideration of SOC in DFT-NEGF calculations, we could determine $T^{\mathrm{M,PMA}}_\perp$ by $T^{\mathrm{S,XC}}_\perp$.

From Fig.~\ref{fig_caseC} (b) we can find that the angle $\theta$ dependence of the cumulative $T^{\mathrm{S,XC/SO}}$ are in the form of $\sin2\theta$ as suggested in Eq.~(\ref{eq_t_pma}), which implies that $\vect{M}$ exhibits a stable state with lowest energy at $\theta = 0^{\circ}$ and $180^{\circ}$ but meta-stable state with highest energies at $\theta=90^{\circ}$.
The magnetic moments aligned along these angles give zero spin torque at these stable/meta-stable states, and experience a torque out of those angles.
The strong positive (negative) $T^{\mathrm{S,XC}}_\perp$ tends to align $\vec{M}$ along the $+z$ ($-z$) direction, as demonstrated by the schematics of Fig.~\ref{fig_caseC}(b).

Moreover, the largest magnitude of $\abs{T^{\mathrm{S,XC}}} \approx \abs{T^{\mathrm{S,SO}}}$ is about $0.34$ meV along with the much smaller $\abs{\Phi^{\mathrm{S,KE}}} \approx \abs{\Phi^{\mathrm{S,SO}}}$ in the order of $\mu$eV.
Comparing to the very weak interlayer exchange coupling between the two Fe electrodes, i.e., $T^{\mathrm{IEC}}$ of case A is in the order of $0.1$ $\mu$eV, it demonstrates that the interfacial SOC dominates the main contribution of the moment torque, i.e,
\begin{equation}
	T^{\mathrm{M,QM}}_\perp = T^{\mathrm{S,XC}}_\perp \approx -T^{\mathrm{S,SO}}_\perp
\end{equation}
from Eq.~(\ref{Tnet}).
Eq.~(\ref{eq_t_pma}) gives that $K_{1}^{\mathrm{PMA}}$ can be estimated by the calculated $T^{\mathrm{S,XC}}_\perp(\theta=45^{\circ})=0.7$ mJ/m$^2$. It is only slightly smaller than the experimental measured PMA of $1 \pm 0.1$ mJ/m$^2$ in the epitaxial Fe/MgO interfaces \cite{Dieny2017} and the Ta/CoFeB/MgO interface \cite{Ikeda2010}, since here we ignore an additional PMA effect from the interface between Fe and heavy metals.

\subsection{Layer-resolved spin torque}

\begin{figure}
	\centering
	\includegraphics{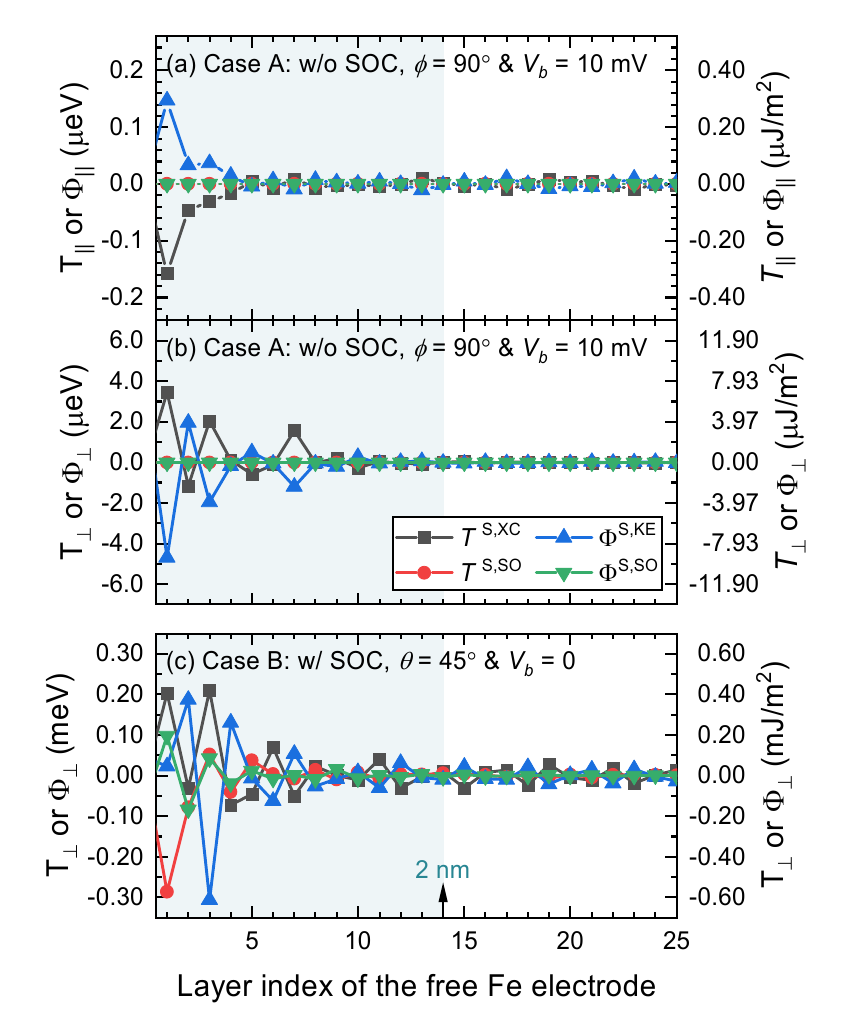}
	\caption{\label{fig_layer}The layer dependency of $T^{\,\mathrm{S,XC/SO}}$ and $\Phi^{\mathrm{S,KE/SO}}$ in the damp-like ($\parallel$) or field-like ($\perp$) direction. (a) and (b) are case A that are without SOC but under an external electrical bias $V_b=10$ mV with an in-plane angle $\phi=90^\circ$. (c) is of case B with an out-of-plane angle $\theta=45^\circ$ in the consideration of SOC at equilibrium ($V_b=0$).}
\end{figure}

In Fig.~\ref{fig_layer}, we present the layer-resolved spin torques and spin current accumulations for case A and B, where the colored region indicates the 2 nm thickness.
The chosen angles meet the maximum values of $T^{\,\mathrm{S,XC/SO}}$ and $\Phi^{\mathrm{S,KE/SO}}$ as shown in the angular dependency of Figs.~\ref{fig_caseA} and \ref{fig_caseC}.
We can observe that the spin torques have only significant influence on the few Fe atoms staring from the MgO/Fe interface.
It reveals that the use of the interfacial effects including the current-driven STT and SOC-induced PMA requires the thin thickness of magnetic films.
Experimentally it has been shown that the thickness of magnetic films is critical to preserve the interfacial effects such as the PMA in Ref.~\cite{Ikeda2010} with the thickness of 1.3 nm.
By our calculations, the penetration and magnitude of the interfacial effects are quantitatively presented via the spin torques and spin current accumulations, which provide a promising way in the analysis of heterojunctions for interfacial spintronics.

\section{Conclusion}


Based on the dynamics of local magnetic moments derived via the spin continuity equation, we propose an exhaustive analytical derivation to calculate spin torques and spin current accumulations.
Two important cases of Fe/MgO/Fe MTJs with noncollinear magnetic configurations have been studied in the framework of first-principles calculations.
Case A is current-driven STT calculation without SOC, which quantitatively agree with previous experiments of MgO-based MTJ with in-plane magnetization along with the linear and quasi-quadratic bias dependence of $T_\parallel$ and $T_\perp$.
In Case B, our DFT-based SOT calculation at equilibrium confirms that the PMA behavior in MgO-based MTJ with out-of-plane magnetization is dominated by the interfacial SOT, which is consistent with experimental PMA of epitaxial Fe/MgO interfaces.
In contract to the conventional DFT-based total energy method, we hereby propose a DFT-based SOT calculation method to quantitatively investigate the underlying mechanism of complex magnetic heterostructures.

\section*{Acknowledgments}

This work was supported by the National Science and Technology Council, Taiwan (NSTC 108-2628-M-008-004-MY3 / 111-2112-M-008-025 / 111-2112-M-001-081 / 110-2923-M-001-003-MY3), the National Center for Theoretical Sciences and the National Center for High- performance Computing for providing computational and storage resources.

\appendix

\section{}
\label{appendix_tmqm}

To study the quantum-mechanical (QM) effect in the influence of magnetic moment dynamics, we make use of the Landau-Lifshitz (LL) equation to involve the effect into an effective magnetic field $\vec{B}_{\lambda}^{\mathrm{eff}}$:
\begin{equation}
	\left. \frac{1}{\gamma} \frac{d\vec{M}_{\lambda}}{dt} \right|_{\mathrm{QM}} = -\vec{M}_{\lambda}\times \vec{B}_{\lambda}^{\mathrm{eff}}~.
\end{equation}
The effective field can be obtained from the exchange-correlation (XC) energy as \cite{Sun2000,Stiles2005,Sharma2007,Eich2013}
\begin{equation}
	\vec{B}_{\lambda}^{\mathrm{eff}}=-\frac{\partial E_{\lambda}^{\mathrm{XC}}}{\partial \vec{M}_{\lambda}}
\end{equation}
with the XC energy computed from the XC Hamiltonian
\begin{equation}
	\begin{split}
		E_{\lambda}^{\mathrm{XC}}&=\langle \hat{H}_{\lambda}^{\mathrm{XC}}\rangle
		\\
		&=\langle \frac{1}{2}\left( \hat{P}_{\lambda}^{\dagger}\hat{H}^{\mathrm{XC}}+\hat{H}^{\mathrm{XC}}\hat{P}_{\lambda} \right) \rangle
		\\
		&=\xi_{\lambda}^{\mathrm{XC}}\langle \frac{1}{2}\left( \hat{P}_{\lambda}^{\dagger}\hatvect{S}+\hatvect{S}\hat{P}_{\lambda} \right) \rangle \cdot \vec{M}_{\lambda}
		\\
		&=\xi_{\lambda}^{\mathrm{XC}} \vec{S}_{\lambda} \cdot \vec{M}_{\lambda}~.
	\end{split}
\end{equation}
Combined the equations above, the QM torque is thus expressed as
\begin{equation}
	\left. \frac{1}{\gamma} \frac{d\vec{M}_{\lambda}}{dt} \right|_{\mathrm{QM}} = \xi_{\lambda}^{\mathrm{XC}}\vec{M}_{\lambda}\times \vec{S}_{\lambda}~.
\end{equation}

\section{Extracting Hamiltonian components}
\label{appendix_h_components}

Given a total (spin-dependent) Hamiltonian matrix from first-principles calculations
\begin{equation}\label{key}
	\mathbf{H} = \mathbf{H}^0 + \mathbf{H}^\mathrm{XC} + \mathbf{H}^\mathrm{SO},
\end{equation}
we seek to find a simple way to separate the spin-independent, exchange and spin-orbit contributions efficiently. For simplicity, we omit the orbital index $\mu$ and $\nu$ in the following derivation since the same operation are applied to each $\mathbf{H}_{\mu\nu}$, which is a $2\times2$ matrix in spin space.

For the spin-independent part $\mathbf{H}^0$, it is approximately obtained by taking the spin-average between the spin-up and -down components \cite{Pareek2002,Theodonis2007,Kalitsov2017}:
\begin{equation}
	\mat{H}^{0}\approx \frac{1}{2}\left( \mat{H}_{\uparrow \uparrow}+\mat{H}_{\downarrow \downarrow} \right) \mat{\sigma }^0,
\end{equation}
where $\mat{\sigma }^0$ is a $2\times2$ identity matrix. For the spin-dependent parts including $\mathbf{H}^\mathrm{XC}$ and $\mathbf{H}^\mathrm{SO}$, it has been shown that \cite{FernndezSeivane2006} the spin Hamiltonian without spin-orbit contribution satisfies the Hermitian property as
\begin{equation}
	\mathbf{H}_{\sigma \sigma '}^{\mathrm{XC}}=\left( \mathbf{H}_{\sigma '\sigma}^{\mathrm{XC}} \right) ^*,
\end{equation}
and the spin-orbit contribution satisfies the anti-Hermitian property as
\begin{equation}\label{key}
	\mathbf{H}_{\sigma \sigma '}^{\mathrm{SO}}=-\left( \mathbf{H}_{\sigma '\sigma}^{\mathrm{SO}} \right) ^*,
\end{equation}
where $\sigma$ and $\sigma'$ denote the index of spin-up or -down. With these properties, the spin Hamiltonian can be uniquely assumed as
\begin{align}
	\mathbf{H}^{\mathrm{XC}} &= \left( \begin{matrix}
		a&		c+id\\
		c-id&		b\\
	\end{matrix} \right)
	\\
	\mathbf{H}^{\mathrm{SO}} &= \left( \begin{matrix}
		i\alpha&		\gamma +i\delta\\
		-\gamma +i\delta&		i\beta\\
	\end{matrix} \right),
\end{align}
where the parameters $\{a,b,c,d\}$ and $\{ \alpha, \beta, \gamma, \delta \}$ are to be determined. Given the total and spin-independent Hamiltonian, the spin-dependent part can be expressed as
\begin{gather}
	\begin{split}
		\mathbf{H}^{\mathrm{S}} &= \mathbf{H}-\mathbf{H}^0 \\
		&=\mathbf{H}^{\mathrm{XC}}+\mathbf{H}^{\mathrm{SO}}
	\end{split} \\
	=\left( \begin{matrix}
		a+i\alpha&		\left( c+\gamma \right) +i\left( d+\delta \right)\\
		\left( c-\gamma \right) -i\left( d-\delta \right)&		b+i\beta\\
	\end{matrix} \right). \notag
\end{gather}
With the expressions
\begin{equation}
	\begin{split}
		\mathbf{H}_{\uparrow \downarrow}^{\mathrm{S}}+\mathbf{H}_{\downarrow \uparrow}^{\mathrm{S}}&=2c+2i\delta
		\\
		\mathbf{H}_{\uparrow \downarrow}^{\mathrm{S}}-\mathbf{H}_{\downarrow \uparrow}^{\mathrm{S}}&=2\gamma +2id~,
	\end{split}
\end{equation}
we can find that the parameters can be determined by
\begin{equation}
	\begin{split}
		a&=\mathrm{Re}\left( \mathbf{H}_{\uparrow \uparrow}^{\mathrm{S}} \right)
		\\
		b&=\mathrm{Re}\left( \mathbf{H}_{\downarrow \downarrow}^{\mathrm{S}} \right)
		\\
		c&=\mathrm{Re}\left( \mathbf{H}_{\uparrow \downarrow}^{\mathrm{S}}+\mathbf{H}_{\downarrow \uparrow}^{\mathrm{S}} \right) /2
		\\
		d&=\mathrm{Im}\left( \mathbf{H}_{\uparrow \downarrow}^{\mathrm{S}}-\mathbf{H}_{\downarrow \uparrow}^{\mathrm{S}} \right) /2
	\end{split}
\end{equation}
and
\begin{equation}
	\begin{split}
		\alpha &=\mathrm{Im}\left( \mathbf{H}_{\uparrow \uparrow}^{\mathrm{S}} \right)
		\\
		\beta &=\mathrm{Im}\left( \mathbf{H}_{\downarrow \downarrow}^{\mathrm{S}} \right)
		\\
		\gamma &=\mathrm{Re}\left( \mathbf{H}_{\uparrow \downarrow}^{\mathrm{S}}-\mathbf{H}_{\downarrow \uparrow}^{\mathrm{S}} \right) /2
		\\
		\delta &=\mathrm{Im}\left( \mathbf{H}_{\uparrow \downarrow}^{\mathrm{S}}+\mathbf{H}_{\downarrow \uparrow}^{\mathrm{S}} \right) /2~.
	\end{split}
\end{equation}

\bibliographystyle{elsarticle-num}
\bibliography{manuscript}

\begin{thebibliography}{10}
\expandafter\ifx\csname url\endcsname\relax
  \def\url#1{\texttt{#1}}\fi
\expandafter\ifx\csname urlprefix\endcsname\relax\def\urlprefix{URL }\fi
\expandafter\ifx\csname href\endcsname\relax
  \def\href#1#2{#2} \def\path#1{#1}\fi

\bibitem{Ikeda2010}
S.~Ikeda, K.~Miura, H.~Yamamoto, K.~Mizunuma, H.~D. Gan, M.~Endo, S.~Kanai,
  J.~Hayakawa, F.~Matsukura, H.~Ohno, \href{https://doi.org/10.1038/nmat2804}{A
  perpendicular-anisotropy {CoFeB}{\textendash}{MgO} magnetic tunnel junction},
  Nat. Mater. 9~(9) (2010) 721--724.
\newline\urlprefix\url{https://doi.org/10.1038/nmat2804}

\bibitem{Weinert1985}
M.~Weinert, R.~E. Watson, J.~W. Davenport,
  \href{https://link.aps.org/doi/10.1103/PhysRevB.32.2115}{Total-energy
  differences and eigenvalue sums}, Phys. Rev. B 32 (1985) 2115--2119.
\newblock \href {https://doi.org/10.1103/PhysRevB.32.2115}
  {\path{doi:10.1103/PhysRevB.32.2115}}.
\newline\urlprefix\url{https://link.aps.org/doi/10.1103/PhysRevB.32.2115}

\bibitem{Daalderop1990}
G.~H.~O. Daalderop, P.~J. Kelly, M.~F.~H. Schuurmans,
  \href{https://link.aps.org/doi/10.1103/PhysRevB.41.11919}{First-principles
  calculation of the magnetocrystalline anisotropy energy of iron, cobalt, and
  nickel}, Phys. Rev. B 41 (1990) 11919--11937.
\newblock \href {https://doi.org/10.1103/PhysRevB.41.11919}
  {\path{doi:10.1103/PhysRevB.41.11919}}.
\newline\urlprefix\url{https://link.aps.org/doi/10.1103/PhysRevB.41.11919}

\bibitem{Peng2015}
S.~Peng, M.~Wang, H.~Yang, L.~Zeng, J.~Nan, J.~Zhou, Y.~Zhang, A.~Hallal,
  M.~Chshiev, K.~L. Wang, Q.~Zhang, W.~Zhao,
  \href{https://doi.org/10.1038/srep18173}{Origin of interfacial perpendicular
  magnetic anisotropy in {MgO}/{CoFe}/metallic capping layer structures}, Sci.
  Rep. 5~(1) (2015) 18173.
\newline\urlprefix\url{https://doi.org/10.1038/srep18173}

\bibitem{Peng2017}
S.~Peng, W.~Zhao, J.~Qiao, L.~Su, J.~Zhou, H.~Yang, Q.~Zhang, Y.~Zhang,
  C.~Grezes, P.~K. Amiri, K.~L. Wang, Giant interfacial perpendicular magnetic
  anisotropy in {MgO}/{CoFe}/capping layer structures, Appl. Phys. Lett.
  110~(7) (2017) 072403.
\newblock \href {https://doi.org/10.1063/1.4976517}
  {\path{doi:10.1063/1.4976517}}.

\bibitem{Wang1996}
X.~Wang, R.~Wu, D.~sheng Wang, A.~J. Freeman, Torque method for the theoretical
  determination of magnetocrystalline anisotropy, Phys. Rev. B 54~(1) (1996)
  61--64.
\newblock \href {https://doi.org/10.1103/physrevb.54.61}
  {\path{doi:10.1103/physrevb.54.61}}.

\bibitem{Pick1999}
{\v{S}}.~Pick,
  \href{https://www.sciencedirect.com/science/article/pii/S0038109899001702}{Magnetic
  anisotropy calculation: implementation of the torque method into the
  recursion-technique scheme}, Solid State Communications 111~(1) (1999)
  15--18.
\newblock \href {https://doi.org/https://doi.org/10.1016/S0038-1098(99)00170-2}
  {\path{doi:https://doi.org/10.1016/S0038-1098(99)00170-2}}.
\newline\urlprefix\url{https://www.sciencedirect.com/science/article/pii/S0038109899001702}

\bibitem{Locatelli2013}
N.~Locatelli, V.~Cros, J.~Grollier,
  \href{https://doi.org/10.1038/nmat3823}{Spin-torque building blocks}, Nature
  Materials 13~(1) (2013) 11--20.
\newblock \href {https://doi.org/10.1038/nmat3823}
  {\path{doi:10.1038/nmat3823}}.
\newline\urlprefix\url{https://doi.org/10.1038/nmat3823}

\bibitem{Manchon2019}
A.Manchon, J.~{\v{Z}}elezn{\'{y}}, I.~Miron, T.~Jungwirth, J.~Sinova,
  A.~Thiaville, K.~Garello, P.~Gambardella,
  \href{https://doi.org/10.1103/revmodphys.91.035004}{Current-induced
  spin-orbit torques in ferromagnetic and antiferromagnetic systems}, Rev. Mod.
  Phys. 91~(3) (Sep. 2019).
\newline\urlprefix\url{https://doi.org/10.1103/revmodphys.91.035004}

\bibitem{JunPy}
The detailed information of \textsc{JunPy} package can be found at the website,
  \url{https://labstt.phy.ncu.edu.tw/junpy/}.

\bibitem{Huang2021}
B.-H. Huang, C.-C. Chao, Y.-H. Tang,
  \href{https://doi.org/10.1063/9.0000117}{Thickness dependence of spin torque
  effect in fe/{MgO}/fe magnetic tunnel junction: Implementation of
  divide-and-conquer with first-principles calculation}, {AIP} Adv. 11~(1)
  (2021) 015036.
\newline\urlprefix\url{https://doi.org/10.1063/9.0000117}

\bibitem{Tang2021}
Y.-H. Tang, B.-H. Huang,
  \href{https://doi.org/10.1103/physrevresearch.3.033264}{Underlying mechanism
  for exchange bias in single-molecule magnetic junctions}, Phys. Rev. Res.
  3~(3) (Sep. 2021).
\newline\urlprefix\url{https://doi.org/10.1103/physrevresearch.3.033264}

\bibitem{Zhang2004}
S.~Zhang, Z.~Li, \href{https://doi.org/10.1103/physrevlett.93.127204}{Roles of
  nonequilibrium conduction electrons on the magnetization dynamics of
  ferromagnets}, Phys. Rev. Lett. 93~(12) (Sep. 2004).
\newline\urlprefix\url{https://doi.org/10.1103/physrevlett.93.127204}

\bibitem{Ralph2008}
D.~Ralph, M.~Stiles, \href{https://doi.org/10.1016/j.jmmm.2007.12.019}{Spin
  transfer torques}, J. Magn. Magn. Mater. 320~(7) (2008) 1190--1216.
\newline\urlprefix\url{https://doi.org/10.1016/j.jmmm.2007.12.019}

\bibitem{Manchon2009}
A.~Manchon, S.~Zhang, \href{https://doi.org/10.1103/physrevb.79.094422}{Theory
  of spin torque due to spin-orbit coupling}, Phys. Rev. B 79~(9) (Mar. 2009).
\newline\urlprefix\url{https://doi.org/10.1103/physrevb.79.094422}

\bibitem{Haney2010}
P.~M. Haney, M.~D. Stiles, {Current-Induced Torques in the Presence of
  Spin-Orbit Coupling}, Phys.l Rev. Lett. 105~(12) (2010) 126602.
\newblock \href {https://doi.org/10.1103/physrevlett.105.126602}
  {\path{doi:10.1103/physrevlett.105.126602}}.

\bibitem{Haney2007}
P.~M. Haney, D.~Waldron, R.~A. Duine, A.~S. N{\'{u}}{\~{n}}ez, H.~Guo, A.~H.
  MacDonald, \href{https://doi.org/10.1103/physrevb.76.024404}{Current-induced
  order parameter dynamics: Microscopic theory applied to {Co/Cu/Co} spin
  valves}, Phys. Rev. B 76~(2) (Jul. 2007).
\newline\urlprefix\url{https://doi.org/10.1103/physrevb.76.024404}

\bibitem{Kalitsov2006}
A.~Kalitsov, I.~Theodonis, N.~Kioussis, M.~Chshiev, W.~H. Butler, A.~Vedyayev,
  \href{https://doi.org/10.1063/1.2151796}{Spin-polarized current-induced
  torque in magnetic tunnel junctions}, J. Appl. Phys. 99~(8) (2006) 08G501.
\newline\urlprefix\url{https://doi.org/10.1063/1.2151796}

\bibitem{Todorov2002}
T.~N. Todorov, \href{https://doi.org/10.1088/0953-8984/14/11/314}{Tight-binding
  simulation of current-carrying nanostructures}, J. Phys.: Condens. Matter
  14~(11) (2002) 3049--3084.
\newline\urlprefix\url{https://doi.org/10.1088/0953-8984/14/11/314}

\bibitem{Tsymbal2019}
E.~Y. Tsymbal, {\v{Z}}.~Igor,
  \href{https://doi.org/10.1201/9780429423079}{Spintronics Handbook: Spin
  Transport and Magnetism, Second Edition}, {CRC} Press, 2019.
\newline\urlprefix\url{https://doi.org/10.1201/9780429423079}

\bibitem{Go2020}
D.~Go, F.~Freimuth, J.-P. Hanke, F.~Xue, O.~Gomonay, K.-J. Lee, S.~Bl\"{u}gel,
  P.~M. Haney, H.-W. Lee, Y.~Mokrousov,
  \href{https://doi.org/10.1103/physrevresearch.2.033401}{Theory of
  current-induced angular momentum transfer dynamics in spin-orbit coupled
  systems}, Phys. Rev. Res. 2~(3) (Sep. 2020).
\newline\urlprefix\url{https://doi.org/10.1103/physrevresearch.2.033401}

\bibitem{Landau1992}
L.~Landau, E.~Lifshitz,
  \href{https://doi.org/10.1016/b978-0-08-036364-6.50008-9}{On the theory of
  the dispersion of magnetic permeability in ferromagnetic bodies}, in:
  Perspectives in Theoretical Physics, Elsevier, 1992, pp. 51--65.
\newline\urlprefix\url{https://doi.org/10.1016/b978-0-08-036364-6.50008-9}

\bibitem{Gilbert2004}
T.~Gilbert, \href{https://doi.org/10.1109/tmag.2004.836740}{Classics in
  magnetics a phenomenological theory of damping in ferromagnetic materials},
  {IEEE} Trans. Magn. 40~(6) (2004) 3443--3449.
\newline\urlprefix\url{https://doi.org/10.1109/tmag.2004.836740}

\bibitem{NL2020}
K.~Dolui, M.~D. Petrović, K.~Zollner, P.~Plecháč, J.~Fabian, B.~K. Nikolić,
  \href{https://doi.org/10.1021/acs.nanolett.9b04556}{Proximity spin–orbit
  torque on a two-dimensional magnet within van der waals heterostructure:
  Current-driven antiferromagnet-to-ferromagnet reversible nonequilibrium phase
  transition in bilayer cri3}, Nano Lett. 20~(4) (2020) 2288--2295.
\newline\urlprefix\url{https://doi.org/10.1021/acs.nanolett.9b04556}

\bibitem{Soriano2014}
M.~Soriano, J.~J. Palacios,
  \href{https://doi.org/10.1103/physrevb.90.075128}{Theory of projections with
  nonorthogonal basis sets: Partitioning techniques and effective
  hamiltonians}, Phys. Rev. B 90~(7) (Aug. 2014).
\newline\urlprefix\url{https://doi.org/10.1103/physrevb.90.075128}

\bibitem{Taylor2001}
J.~Taylor, H.~Guo, J.~Wang,
  \href{https://doi.org/10.1103/physrevb.63.245407}{{Ab initio} modeling of
  quantum transport properties of molecular electronic devices}, Phys. Rev. B
  63~(24) (Jun. 2001).
\newline\urlprefix\url{https://doi.org/10.1103/physrevb.63.245407}

\bibitem{Theodonis2006}
I.~Theodonis, N.~Kioussis, A.~Kalitsov, M.~Chshiev, W.~H. Butler,
  \href{https://doi.org/10.1103/physrevlett.97.237205}{Anomalous bias
  dependence of spin torque in magnetic tunnel junctions}, Phys. Rev. Lett.
  97~(23) (Dec. 2006).
\newline\urlprefix\url{https://doi.org/10.1103/physrevlett.97.237205}

\bibitem{Tang2010}
Y.-H. Tang, N.~Kioussis, A.~Kalitsov, W.~H. Butler, R.~Car,
  \href{https://doi.org/10.1103/physrevb.81.054437}{Influence of asymmetry on
  bias behavior of spin torque}, Phys. Rev. B 81~(5) (Feb. 2010).
\newline\urlprefix\url{https://doi.org/10.1103/physrevb.81.054437}

\bibitem{Datta1995}
S.~Datta, \href{https://doi.org/10.1017/CBO9780511805776}{Electronic Transport
  in Mesoscopic Systems}, Cambridge Studies in Semiconductor Physics and
  Microelectronic Engineering, Cambridge University Press, 1995.
\newline\urlprefix\url{https://doi.org/10.1017/CBO9780511805776}

\bibitem{Brandbyge2002}
M.~Brandbyge, J.-L. Mozos, P.~Ordej{\'{o}}n, J.~Taylor, K.~Stokbro,
  \href{https://doi.org/10.1103/physrevb.65.165401}{Density-functional method
  for nonequilibrium electron transport}, Phys. Rev. B 65~(16) (Mar. 2002).
\newline\urlprefix\url{https://doi.org/10.1103/physrevb.65.165401}

\bibitem{Kresse1996}
G.~Kresse, J.~Furthm\"{u}ller,
  \href{https://doi.org/10.1103/physrevb.54.11169}{Efficient iterative schemes
  for {ab inition} total-energy calculations using a plane-wave basis set},
  Phys. Rev. B 54~(16) (1996) 11169--11186.
\newline\urlprefix\url{https://doi.org/10.1103/physrevb.54.11169}

\bibitem{Kresse1999}
G.~Kresse, D.~Joubert, \href{https://doi.org/10.1103/physrevb.59.1758}{From
  ultrasoft pseudopotentials to the projector augmented-wave method}, Phys.
  Rev. B 59~(3) (1999) 1758--1775.
\newline\urlprefix\url{https://doi.org/10.1103/physrevb.59.1758}

\bibitem{Perdew1996}
J.~P. Perdew, K.~Burke, M.~Ernzerhof,
  \href{https://doi.org/10.1103/physrevlett.77.3865}{Generalized gradient
  approximation made simple}, Phys. Rev. Lett. 77~(18) (1996) 3865--3868.
\newline\urlprefix\url{https://doi.org/10.1103/physrevlett.77.3865}

\bibitem{Waldron2007}
D.~Waldron, L.~Liu, H.~Guo,
  \href{https://doi.org/10.1088/0957-4484/18/42/424026}{{Ab initio} simulation
  of magnetic tunnel junctions}, Nanotechnology 18~(42) (2007) 424026.
\newline\urlprefix\url{https://doi.org/10.1088/0957-4484/18/42/424026}

\bibitem{Haney2008}
P.~Haney, R.~Duine, A.~N{\'{u}}{\~{n}}ez, A.~MacDonald,
  \href{https://doi.org/10.1016/j.jmmm.2007.12.020}{Current-induced torques in
  magnetic metals: Beyond spin-transfer}, J. Magn. Magn. Mater. 320~(7) (2008)
  1300--1311.
\newline\urlprefix\url{https://doi.org/10.1016/j.jmmm.2007.12.020}

\bibitem{Tang2018}
Y.-H. Tang, B.-H. Huang,
  \href{https://doi.org/10.1021/acs.jpcc.8b03772}{Manipulation of giant
  field-like spin torque in amine-ended single-molecule magnetic junctions}, J.
  Phys. Chem. C 122~(35) (2018) 20500--20505.
\newline\urlprefix\url{https://doi.org/10.1021/acs.jpcc.8b03772}

\bibitem{Lee2015}
K.-S. Lee, D.~Go, A.~Manchon, P.~M. Haney, M.~D. Stiles, H.-W. Lee, K.-J. Lee,
  \href{https://doi.org/10.1103/physrevb.91.144401}{Angular dependence of
  spin-orbit spin-transfer torques}, Phys. Rev. B 91~(14) (Apr. 2015).
\newline\urlprefix\url{https://doi.org/10.1103/physrevb.91.144401}

\bibitem{Stile2008}
C.~Heiliger, M.~D. Stiles,
  \href{https://link.aps.org/doi/10.1103/PhysRevLett.100.186805}{Ab initio
  studies of the spin-transfer torque in magnetic tunnel junctions}, Phys. Rev.
  Lett. 100 (2008) 186805.
\newline\urlprefix\url{https://link.aps.org/doi/10.1103/PhysRevLett.100.186805}

\bibitem{Guo2011}
X.~Jia, K.~Xia, Y.~Ke, H.~Guo,
  \href{https://link.aps.org/doi/10.1103/PhysRevB.84.014401}{Nonlinear bias
  dependence of spin-transfer torque from atomic first principles}, Phys. Rev.
  B 84 (2011) 014401.
\newline\urlprefix\url{https://link.aps.org/doi/10.1103/PhysRevB.84.014401}

\bibitem{Kubota2008}
H.~Kubota, A.~Fukushima, K.~Yakushiji, T.~Nagahama, S.~Yuasa, K.~Ando,
  H.~Maehara, Y.~Nagamine, K.~Tsunekawa, D.~D. Djayaprawira, N.~Watanabe,
  Y.~Suzuki, Quantitative measurement of voltage dependence of spin-transfer
  torque in mgo-based magnetic tunnel junctions, Nat. Phys. 4~(1) (2008)
  37--41.
\newblock \href {https://doi.org/10.1038/nphys784}
  {\path{doi:10.1038/nphys784}}.

\bibitem{Sankey2008}
J.~C. Sankey, Y.-T. Cui, J.~Z. Sun, J.~C. Slonczewski, R.~A. Buhrman, D.~C.
  Ralph, Measurement of the spin-transfer-torque vector in magnetic tunnel
  junctions, Nat. Phys. 4~(1) (2008) 67--71.
\newblock \href {https://doi.org/10.1038/nphys783}
  {\path{doi:10.1038/nphys783}}.

\bibitem{Stiles2002}
M.~D. Stiles, A.~Zangwill,
  \href{https://doi.org/10.1103/physrevb.66.014407}{Anatomy of spin-transfer
  torque}, Phys. Rev. B 66~(1) (Jun. 2002).
\newline\urlprefix\url{https://doi.org/10.1103/physrevb.66.014407}

\bibitem{Slonczewski2005}
J.~C. Slonczewski, \href{https://doi.org/10.1103/physrevb.71.024411}{Currents,
  torques, and polarization factors in magnetic tunnel junctions}, Phys. Rev. B
  71~(2) (Jan. 2005).
\newline\urlprefix\url{https://doi.org/10.1103/physrevb.71.024411}

\bibitem{Tang2009}
Y.-H. Tang, N.~Kioussis, A.~Kalitsov, W.~H. Butler, R.~Car,
  \href{https://link.aps.org/doi/10.1103/PhysRevLett.103.057206}{Controlling
  the nonequilibrium interlayer exchange coupling in asymmetric magnetic tunnel
  junctions}, Phys. Rev. Lett. 103 (2009) 057206.
\newline\urlprefix\url{https://link.aps.org/doi/10.1103/PhysRevLett.103.057206}

\bibitem{Slonczewski1993}
J.~C. Slonczewski,
  \href{https://doi.org/10.1016/0304-8853(93)90630-k}{{Mechanism of interlayer
  exchange in magnetic multilayers}}, J. Magn. Magn. Mater. 126~(1-3) (1993)
  374--379.
\newline\urlprefix\url{https://doi.org/10.1016/0304-8853(93)90630-k}

\bibitem{Bruno1995}
P.~Bruno, \href{https://link.aps.org/doi/10.1103/PhysRevB.52.411}{Theory of
  interlayer magnetic coupling}, Phys. Rev. B 52 (1995) 411--439.
\newblock \href {https://doi.org/10.1103/PhysRevB.52.411}
  {\path{doi:10.1103/PhysRevB.52.411}}.
\newline\urlprefix\url{https://link.aps.org/doi/10.1103/PhysRevB.52.411}

\bibitem{YangPRB2011}
H.~X. Yang, M.~Chshiev, B.~Dieny, J.~H. Lee, A.~Manchon, K.~H. Shin,
  \href{https://link.aps.org/doi/10.1103/PhysRevB.84.054401}{First-principles
  investigation of the very large perpendicular magnetic anisotropy at fe$|$mgo
  and co$|$mgo interfaces}, Phys. Rev. B 84 (2011) 054401.
\newblock \href {https://doi.org/10.1103/PhysRevB.84.054401}
  {\path{doi:10.1103/PhysRevB.84.054401}}.
\newline\urlprefix\url{https://link.aps.org/doi/10.1103/PhysRevB.84.054401}

\bibitem{Dieny2017}
B.~Dieny, M.~Chshiev,
  \href{https://doi.org/10.1103/revmodphys.89.025008}{{Perpendicular magnetic
  anisotropy at transition metal/oxide interfaces and applications}}, Rev. Mod.
  Phys. 89~(2) (2017) 025008.
\newline\urlprefix\url{https://doi.org/10.1103/revmodphys.89.025008}

\bibitem{Sun2000}
J.~Z. Sun, {Spin-current interaction with a monodomain magnetic body: A model
  study}, Phys. Rev. B 62~(1) (2000) 570--578.
\newblock \href {https://doi.org/10.1103/physrevb.62.570}
  {\path{doi:10.1103/physrevb.62.570}}.

\bibitem{Stiles2005}
J.~Xiao, A.~Zangwill, M.~D. Stiles, {Macrospin models of spin transfer
  dynamics}, Phys. Rev. B 72~(1) (2005) 014446.
\newblock \href {https://doi.org/10.1103/physrevb.72.014446}
  {\path{doi:10.1103/physrevb.72.014446}}.

\bibitem{Sharma2007}
S.~Sharma, J.~K. Dewhurst, C.~Ambrosch-Draxl, S.~Kurth, N.~Helbig, S.~Pittalis,
  S.~Shallcross, L.~Nordstr\"om, E.~K.~U. Gross,
  \href{https://link.aps.org/doi/10.1103/PhysRevLett.98.196405}{First-principles
  approach to noncollinear magnetism: Towards spin dynamics}, Phys. Rev. Lett.
  98 (2007) 196405.
\newblock \href {https://doi.org/10.1103/PhysRevLett.98.196405}
  {\path{doi:10.1103/PhysRevLett.98.196405}}.
\newline\urlprefix\url{https://link.aps.org/doi/10.1103/PhysRevLett.98.196405}

\bibitem{Eich2013}
F.~G. Eich, E.~K.~U. Gross,
  \href{https://link.aps.org/doi/10.1103/PhysRevLett.111.156401}{Transverse
  spin-gradient functional for noncollinear spin-density-functional theory},
  Phys. Rev. Lett. 111 (2013) 156401.
\newblock \href {https://doi.org/10.1103/PhysRevLett.111.156401}
  {\path{doi:10.1103/PhysRevLett.111.156401}}.
\newline\urlprefix\url{https://link.aps.org/doi/10.1103/PhysRevLett.111.156401}

\bibitem{Pareek2002}
T.~P. Pareek, P.~Bruno, \href{https://doi.org/10.1103/physrevb.65.241305}{Spin
  coherence in a two-dimensional electron gas with rashba spin-orbit
  interaction}, Phys. Rev. B 65~(24) (May 2002).
\newline\urlprefix\url{https://doi.org/10.1103/physrevb.65.241305}

\bibitem{Theodonis2007}
I.~Theodonis, A.~Kalitsov, N.~Kioussis,
  \href{https://doi.org/10.1103/physrevb.76.224406}{Enhancing spin-transfer
  torque through the proximity of quantum well states}, Phys. Rev. B 76~(22)
  (Dec. 2007).
\newline\urlprefix\url{https://doi.org/10.1103/physrevb.76.224406}

\bibitem{Kalitsov2017}
A.~Kalitsov, S.~A. Nikolaev, J.~Velev, M.~Chshiev, O.~Mryasov,
  \href{https://doi.org/10.1103/physrevb.96.214430}{Intrinsic spin-orbit torque
  in a single-domain nanomagnet}, Phys. Rev. B 96~(21) (Dec. 2017).
\newline\urlprefix\url{https://doi.org/10.1103/physrevb.96.214430}

\bibitem{FernndezSeivane2006}
L.~Fern{\'{a}}ndez-Seivane, M.~A. Oliveira, S.~Sanvito, J.~Ferrer,
  \href{https://doi.org/10.1088/0953-8984/18/34/012}{On-site approximation for
  spin{\textendash}orbit coupling in linear combination of atomic orbitals
  density functional methods}, J. Phys.: Condens. Matter 18~(34) (2006)
  7999--8013.
\newline\urlprefix\url{https://doi.org/10.1088/0953-8984/18/34/012}

\end{thebibliography}

\end{document}